\documentclass[10pt,journal]{IEEEtran}
\usepackage{lipsum} 
\usepackage[tight,footnotesize]{subfigure}
\usepackage{enumitem}
\usepackage{float}
\usepackage{multicol,multienum}
\usepackage{breqn}
\usepackage{stfloats}
\usepackage{multirow}
\usepackage{diagbox}
\usepackage{verbatim}
\usepackage{hyperref}
\usepackage{bm}
\usepackage{cite}
\usepackage{graphicx}
\usepackage{tabularx}
\usepackage{booktabs}
\usepackage{amsthm}
\usepackage{amssymb}
\usepackage{thmtools}

\newtheorem{definition}{Definition}

\usepackage[tight,footnotesize]{subfigure}
\usepackage{url}
\usepackage{doi}
\usepackage{color}
\usepackage{algorithm}
\usepackage{algorithmic}
\usepackage{subfigure}
\usepackage{multirow}
\newenvironment{sloppypar*}
{\sloppy\ignorespaces}
{\par}

\ifCLASSINFOpdf

\else

\fi

\begin{document}\sloppy

\title{Privacy-preserving Anomaly Detection in Cloud Manufacturing via Federated Transformer}

\author{Shiyao Ma, Jiangtian Nie, Jiawen Kang, Lingjuan Lyu, Ryan Wen Liu, Ruihui Zhao, Ziyao Liu, Dusit Niyato,~\IEEEmembership{Fellow,~IEEE}

\thanks{S. Ma is with the College of Information and Communication Engineering, Dalian Minzu Unversity, Dalian, 116600, China (e-mail: shiyaoma.cs@gmail.com). J. Nie is with School of Computer Science and Engineering, Nanyang Technological University, Singapore (e-mail: jnie001@e.ntu.edu.sg). J. Kang is with School of Automation, Guangdong University of Technology, Guangzhou, China (e-mail: kavinkang@gdut.edu.cn). L.~Lyu is with Sony AI, Tokyo, Japan (email: lingjuanlvsmile@gmail.com). R. W. Liu is with the School of Navigation, Wuhan University of Technology, Wuhan 430063, China, and also with the National Engineering Research Center for Water Transport Safety, Wuhan 430063, China (e-mail: wenliu@whut.edu.cn). R. Zhao is with Tencent Jarvis Lab, Shenzhen, China (e-mail: zacharyzhao@tencent.com). Z. Liu is with School of Computer Science and Engineering, Nanyang Technological University, Singapore (e-mail: ziyao002@e.ntu.edu.sg). D. Niyato is with School of Computer Science and Engineering (SCSE), Nanyang Technological University, Singapore (e-mail: dniyato@ntu.edu.sg).  Corresponding author: J. Nie.}}

\markboth{IEEE Transactions on Industrial Informatics}%
{Shell \MakeLowercase{\textit{et al.}}: Bare Demo of IEEEtran.cls for IEEE Journals}

\maketitle

\begin{abstract}
With the rapid development of cloud manufacturing, industrial production with edge computing as the core architecture has been greatly developed. However, edge devices often suffer from abnormalities and failures in industrial production. Therefore, detecting these abnormal situations timely and accurately is crucial for cloud manufacturing. As such, a straightforward solution is that the edge device uploads the data to the cloud for anomaly detection. However, Industry 4.0 puts forward higher requirements for data privacy and security so that it is unrealistic to upload data from edge devices directly to the cloud. Considering the above-mentioned severe challenges, this paper customizes a weakly-supervised edge computing anomaly detection framework, \emph{i.e.}, Federated Learning-based Transformer framework (\textit{FedAnomaly}), to deal with the anomaly detection problem in cloud manufacturing. Specifically, we introduce federated learning (FL) framework that allows edge devices to train an anomaly detection model in collaboration with the cloud without compromising privacy. To boost the privacy performance of the framework, we add differential privacy noise to the uploaded features. To further improve the ability of edge devices to extract abnormal features, we use the Transformer to extract the feature representation of abnormal data. In this context, we design a novel collaborative learning protocol to promote efficient collaboration between FL and Transformer. Furthermore, extensive case studies on four benchmark data sets verify the effectiveness of the proposed framework. To the best of our knowledge, this is the first time integrating FL and Transformer to deal with anomaly detection problems in cloud manufacturing.
\end{abstract}


\IEEEpeerreviewmaketitle

\section{Introduction}
In recent years, the rapid development of information technologies and artificial intelligence technologies have brought about a huge revolution in Industry 4.0 \cite{liu2020deep}, especially the organic integration of cloud manufacturing and big data technology \cite{wan2017manufacturing}. To provide more ``fresh blood'' for industrial manufacturing, many advanced manufacturing systems have been proposed, such as intelligent factory systems and digital twin systems \cite{tao2018digital}. Edge computing \cite{shi2016edge}, as the core architecture of these advanced systems, has attracted the attention of academia and industry. The reason is that edge computing has the inherent advantages of flexible architecture, common standards and specifications, effective operating mechanisms, and collaboration capabilities. Therefore, the edge computing architecture has become the most widely used system in cloud manufacturing and the Industrial Internet of Things (IIoT) and has played an essential role in many industrial plants and factories \cite{6742605}.

Edge computing, as a practical distributed computing paradigm, improves the efficiency of industrial manufacturing by offloading heavy computing tasks from clouds to edge nodes. Specifically, edge computing allows the computing of downstream data representing cloud services and upstream data representing IoT services at the edge of the network \cite{shi2016edge,liu2020privacy}. Furthermore, the organic integration of edge computing and mobile computing has spawned a mobile edge computing (MEC) paradigm, which has further promoted the development of cloud manufacturing in Industry 4.0. Therefore, MEC is becoming the main driving force of the manufacturing industry, promoting the development of Industry 4.0 \cite{7883994}, and has been used in a wide range of fields, such as smart cities and industrial production \cite{liu2021towards,7883994,6742575,liu2020federated}.

Even though MEC has achieved great success in cloud manufacturing, it still suffers from abnormalities (or failures) \cite{liu2020deep}. Abnormalities, \emph{i.e.}, outliers, and out-of-distribution values, generally refer to things that deviate from the standard, normal, or expected. In the context of MEC, exceptions refer typically to equipment or edge devices in cloud manufacturing that fail or generate abnormal values and trigger system alarms \cite{hussain2019artificial}. If abnormalities or malfunctions are not accurately detected, these error cases generally damage the machine and cause economic losses to the factory \cite{7840777}. Therefore, we need efficient detection techniques to detect anomalies in the MEC architecture to protect or maintain edge devices' safety or normal operation (or edge machines). However, an unresolved issue is how to deal with abnormal or unexpected abnormal situations in production in a real-time manner. This indicates that we need to build a bridge to connect MEC and anomaly detection techniques to achieve anomaly detection with timely feedback.

To detect anomalies or failures in MEC in time, many anomaly detection methods have been developed \cite{keshk2019integrated, erfani2014privacy, vasilomanolakis2015taxonomy}, among which these methods include three categories: supervised, unsupervised, and weakly-supervised anomaly detection categories. For example, in \cite{7430287}, the authors proposed a supervised neural network, which detects abnormal devices through the hardware information of the devices. However, the supervised method relies heavily on the labeled data and thus is limited by the precious labeled data and cannot be widely used \cite{liu2020rc}. To address this issue, in \cite{9208761}, the authors proposed a unsupervised GRU-based VAE model to detect the anomalies in IIoT systems. However, its performance is still very different from supervised anomaly detection methods. Therefore, researchers focus on weakly-supervised anomaly detection methods, which can utilize only a small amount of anomaly data to achieve efficient anomaly detection so that the model's performance is close to supervised anomaly detection. For example, in \cite{liu2020deep}, the authors proposed a weakly-supervised deep learning-based optimization method for anomalies detection in IIoT. 

Although the existing anomaly detection methods that are weakly-supervised methods have achieved good results in the MEC architecture, there are still the following issues: 1) \textbf{Privacy Leakage:} The previous techniques are directly uploading data to the cloud for anomaly detection, which not only leads to users' privacy leaked, but also increases the communication overhead. 2) \textbf{Poor Scalability:} The previous methods cannot be updated efficiently or require expensive costs to update the on-device anomaly detection model. Indeed, with the improvement of data privacy protection laws and regulations such as GDPR \cite{voigt2017eu}, central and remote processing of raw data of edge devices in the cloud is not possible, and the data cannot be shared between edge nodes. Therefore, we need to design a privacy-protected and extensible MEC-based anomaly detection framework to detect anomalies.

In this paper, we propose a federated learning-based Transformer framework for anomaly detection in cloud manufacturing, called the FedAnomaly framework, which can efficiently detect anomalies in cloud manufacturing while protecting privacy. In this framework, FL allows distributed edge devices to collaboratively train a shared global model without exchanging data, so that specific tasks can be completed while protecting data privacy. Furthermore, to boost the privacy performance of the proposed framework, we add differential privacy~\cite{dwork2008differential} noise to the features that are produced by Transformer. Nevertheless, the training method of the Transformer conflicts with the training protocol of FL, it leads that is difficult for us to apply the Transformer to FL directly. To solve this issue, we propose a novel FL training protocol to adapt Transformer training. Specifically, in our training protocol, each edge device holds a local encoder, \emph{i.e.}, Transformer, which can extract important abnormal data feature representations and upload these encoded features {with differential privacy noise} to the cloud, and the cloud holds a decoder model, \emph{i.e.}, MLP, can distinguish whether the features uploaded by the edge devices are normal or abnormal. We evaluate our model on four data sets, and the experimental results show that our model exceeds the current advanced anomaly detection model \cite{9465358}. The main contributions of the paper are as follows:



\begin{itemize}
    \item We propose a new federated learning-based Transformer framework, called FedAnomaly, to achieve privacy-preserving anomaly detection in cloud manufacturing in a timely manner. Specifically, on the edge-side, we design a weakly-supervised anomaly feature extraction model based on Transformer, and we add differential privacy noise into the features before the edge devices upload the features. On the cloud-side, we use MLP as our anomaly detection model.
    
    \item We tailor a novel FL training protocol for FedAnomaly, which is no longer exchanging model updates but exchanging encoded features and losses. To this end, the designed training protocol is adapted to Transformer training and is superior to the traditional FL training protocol in terms of communication overhead.
    
    \item We evaluate the effectiveness of FedAnomaly on four benchmark data sets. The experimental results show that the proposed model is better than the advanced anomaly detection models. In addition, we explore the sensitivity of the model to different hyperparameters.
\end{itemize}

\section{Preliminaries}
\subsection{Federated Learning}
Federated Learning \cite{mcmahan2017communication, zheng2022aggregation} is a privacy-protected distributed learning paradigm, which can conduct collaborative training under the premise of protecting participant data privacy. FL involves two entities: cloud and edge devices, where the edge device is responsible for training the local model on its local data set, and the cloud as the coordinator is responsible for aggregating and updating the model. Specifically, in each round of training, the edge device first conducts a local model that is trained by the local data set and then uploads the obtained model to the cloud. The cloud uses an aggregation algorithm (such as FedAvg\cite{mcmahan2017communication}) to aggregate the collected local model and sends the model aggregated to the edge devices. FL repeatedly executes the above training protocol to obtain a convergent global model. Therefore, the goal of the cloud is to minimize the objective function as follows: 
\begin{equation}
\min_w F(w), \text{where }  F(w)= \sum_{i=1}^N p_i F_i(w),    
\end{equation}
where $N$ is the number of the edge devices, $p_i\geq0$ is the weight of the $i$-th edge device during aggregation (where $\sum_{i=1}^N p_i= 1$), and $F_i(w)$ is the $i$-th local objective function.

\subsection{Anomaly Detection}
Anomaly detection is a technique used to detect samples, events, and observations that are very different from certain concepts of normality. The focus of anomaly detection is to detect abnormal samples hidden in normal samples through modeling.  Anomalies are defined as follows:
\begin{definition}(Anomalies).
Given a data set $\mathcal{X} \subseteq \mathbb{R}^d$ contains normal data and abnormal data. The distribution of normal data in $\mathcal{X}$ is $\mathbb{P }$, which represents the true rules of normal behavior in the task. We assume that $\mathbb{P}$ has a probability density function $p$, we have:
\begin{equation}
    \mathrm{Anomalies} = \{x \in \mathcal{X} \mid p(x) \leq \tau\}, \tau \geq 0,
\end{equation}
where $\tau$ is the threshold that can clearly distinguish abnormal from normal and $\tau$ is relatively small, because abnormalities are very rare in the real world.
\end{definition}
Anomaly detection aims to find the distribution of anomalies $\mathbb{P}^-$ through some methods such as probability-based methods \cite{liu2012unsupervised} and neural network (NN)-based methods \cite{9465358}. In this paper, we use anomaly scores based on neural networks to distinguish anomalies, which is formalized as:
\begin{equation}
   \mathrm{Anomalies} = \{ x \in \mathcal{X} \mid f(x) \text{ is closed to $\mathcal{S}$ } \},
\end{equation}
where $f$ is the anomaly detection model and $\mathcal{S}$ is the score of labeled anomaly.

\subsection{Differential Privacy}
Differential privacy (DP) techniques \cite{dwork2008differential} have emerged as the substantial standard for data privacy in recent years. DP assures individuals that no matter what the prior knowledge, \emph{i.e.}, it not leakage their data privacy, and its possibility of participating in data collection can be ignored. Therefore, DP is often used in a distributed environment to protect privacy. The definition of original DP is as follows:
\begin{definition}
($\epsilon$-Differential Privacy). Given a randomized algorithm $\mathcal{Z}$ meets the condition that is $\epsilon$-DP and any data sets $D_1$ and $D_2$, if there is only a different record between $D_1$ and $D_2$, given any set of outputs $\mathcal{O}$, we have:
\begin{equation}
    \Pr[\mathcal{Z}(D_1)\in \mathcal{O}] \leq e^{\epsilon} \cdot \Pr[\mathcal{Z}(D_2) \in \mathcal{O}],
\end{equation}
where $\Pr[\cdot]$ is probability distribution function.
\end{definition}
Later, for convenience, the definition of easy-to-use DP as follows: 
\begin{definition}
(($\epsilon$, $\delta$)-Differential Privacy). Given a randomized algorithm $\mathcal{Z}$ meets the condition that is ($\epsilon$, $\delta$)-DP and any data sets $D_1$ and $D_2$, for any set of outputs $\mathcal{O}$, we have:
\begin{equation}
    \Pr[\mathcal{Z}(D_1)\in \mathcal{O}] \leq e^{\epsilon} \cdot \Pr[\mathcal{Z}(D_2) \in \mathcal{O}]+ \delta,
    \label{eq:differential privacy}
\end{equation}
where $\delta$ is the failure probability, in order to obtain a stable privacy protection effect, we usually set $\delta \in [0,1)$. 
\end{definition}
In practice, differential privacy noise (DP noise) is added to the original data to protect data privacy. Furthermore, we usually utilize the Gaussian noise as the added DP noise. The definition of Gaussian noise is as follows:
\begin{definition}
(Gaussian noise). Given the $f$ is the function that returns the real-value, $\mathcal{N}(0, \Delta f^2 \cdot \sigma^2)$ is the Gaussian  (Normal) distribution and $\Delta f^2$ is the sensitivity of $f$ that can be calculated by $\Delta f = \max \limits_{D_1,D_2} \Vert f(D_1) - f(D_2) \Vert_2$, we have:
\begin{equation}
    \mathcal{Z}(D_1) \triangleq f(D_1) + \mathcal{N}(0, \Delta f^2 \cdot \sigma^2).
    \label{eq:differential_privacy_noise}
\end{equation}
\end{definition}
Finally, we obtain the new data set that adds Gaussian noise and uploads the new data set to the server without leaking the user privacy.

\subsection{Transformer}
\begin{figure}[t]
	\centering
	\includegraphics[width=1\linewidth]{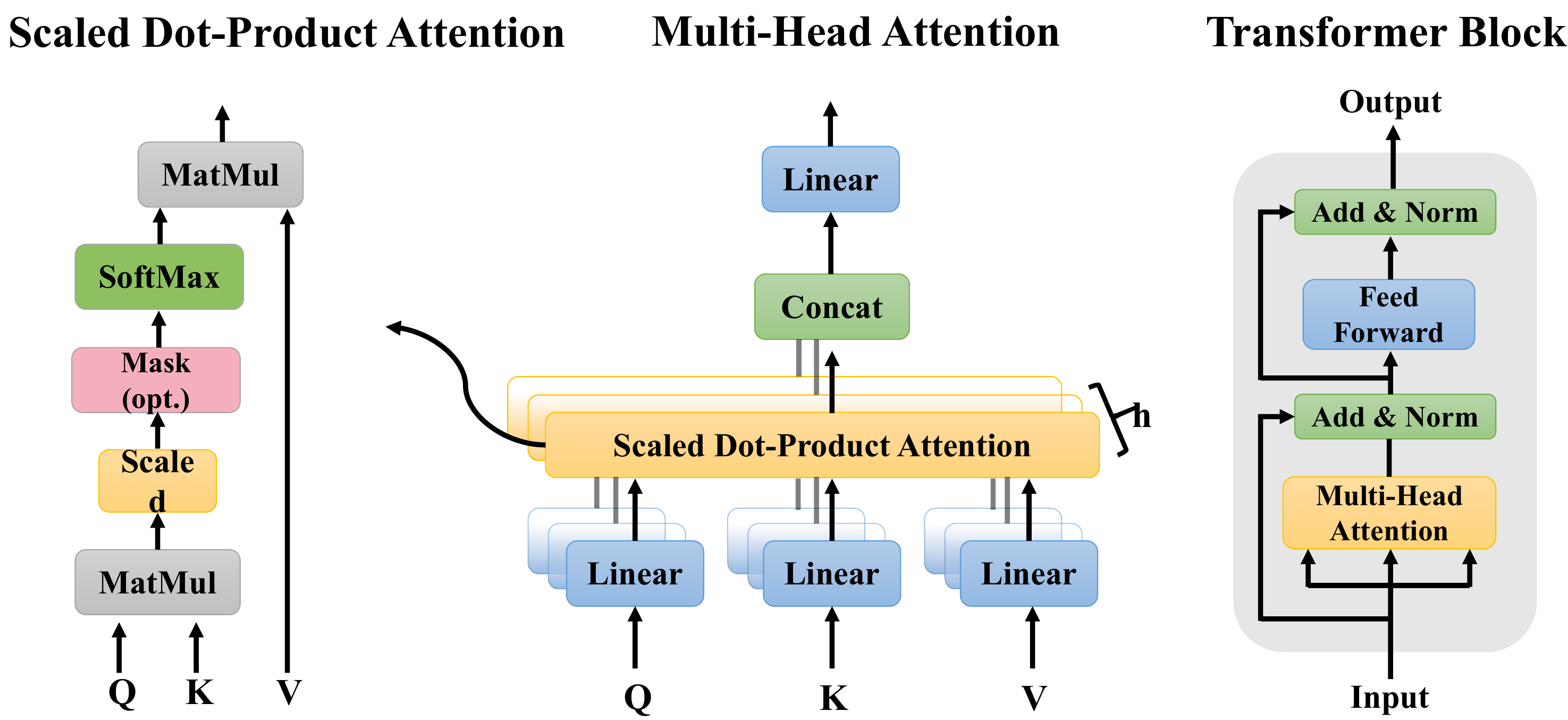}
	\caption{Scaled dot-product attention block (left), multi-head attention block (middle) and Transformer block (right).}
	\label{fig:transformer}
\end{figure}
Transformer is currently the most popular neural network that is based on the self-attention mechanism, which is widely used in AI applications and has achieved a good result. Transformer has two parts: encoder and decoder, in which encoder is responsible for capturing the intrinsic features, and decoder is responsible for decoding the information that we need. Both the encoder and decoder consist of the multi-head attention block, the attention mechanism is the critical component that exists in the multi-head attention block, \emph{i.e.}, as shown in middle of Fig.\ref{fig:transformer}, the multi-head attention block and feed-forward layer make up the transformer block that can be stacked on the top of each other multiples times in encoder and decoder. Significantly, Transformer adopts a residual connection of the two sub-layers and then implements layer normalization. Next, we introduce the training process of Transformer. Firstly, the Transformer structure is different from the recurrent network structure, and Transformer can not remember how the sequence was fed into the model. Therefore, the input and output should be added information that contains the positional information of the elements in the input and output by positional encoding part, which can be expressed as:
\begin{equation}
\begin{aligned}
    \mathrm{Positional\_Encoding}{(p,2i)} = sin(p \cdot \omega_{2i})\\
    \mathrm{Positional\_Encoding}{(p,2i+1)} = cos(p \cdot \omega_{2i}), 
\end{aligned}
\end{equation}
where $p$ is the position, $i$ is the current dimension, $\omega_{2i}=\frac{1}{10000^{2i/d}}$ and $d$ is the data dimension. That is, each dimension has the position information. 
After obtaining the position information, we can utilize multi-head attention block to train the data, it can be expressed as:
\begin{equation}
\begin{aligned}
    \mathrm{Attention}(Q^{(i)},K^{(i)},V^{(i)}) =  \mathrm{softmax}(\frac{Q^{(i)}K^{(i)T}}{\sqrt{d_{k}}})V^{(i)},
    \label{attention}
\end{aligned}
\end{equation}
where $i$ represents the $i$-th head of multi-head attention block, $d_k$ represents the dimension of input data, and $Q^{(i)}$, $K^{(i)}$ and $V^{(i)}$ are the queries, keys, and values of the $i$-th head, respectively. Furthermore, $Q$ is the query matrix that means the input subset, $K$ is the key matrix that means the abstract attribute of the all input data and $V$ is the value matrix that means the complete set of data. In Eq.\eqref{attention}, we firstly dot-product $Q$ and $K^T$ to obtain the similarity between the input sequence and abstract attribute of the all input data, then we utilize $\mathrm{softmax}(\frac{Q^{(i)}K^{(i)T}}{\sqrt{d_{k}}})$ to obtain the similarity weight, we finally dot-product $\mathrm{softmax}(\frac{Q^{(i)}K^{(i)T}}{\sqrt{d_{k}}})$ and $V$ to get the attention score. Furthermore, $Q^{(i)}$, $K^{(i)}$ and $V^{(i)}$ are obtained by the following equation:
\begin{equation}
\left\{ {\begin{array}{*{20}{l}}
  {{Q^{(i)}} = XW_i^Q} \\ 
  {{K^{(i)}} = XW_i^K} \\ 
  {{V^{(i)}} = XW_i^V} 
\end{array},} \right.
\end{equation}
where $W^Q_i$, $W^K_i$ and $W^V_i$ are the weight matrix and $X$ is the input samples. The multi-head attention value is calculated by the following equation:
\begin{equation}
    \mathrm{Multi\_Head\_Attention}(Q,K,V)=\mathrm{Concat}(h_1,...,h_n)W^0,
\end{equation}
where $h_i$ is the value computed by Eq. \eqref{attention} and $W^0$ is the learnable weight matrix to produce the output of the block. Finally, the multi-head attention value is passed through the feed forward layer to get the output.

\begin{figure}[t]
	\centering
	\includegraphics[width=1\linewidth]{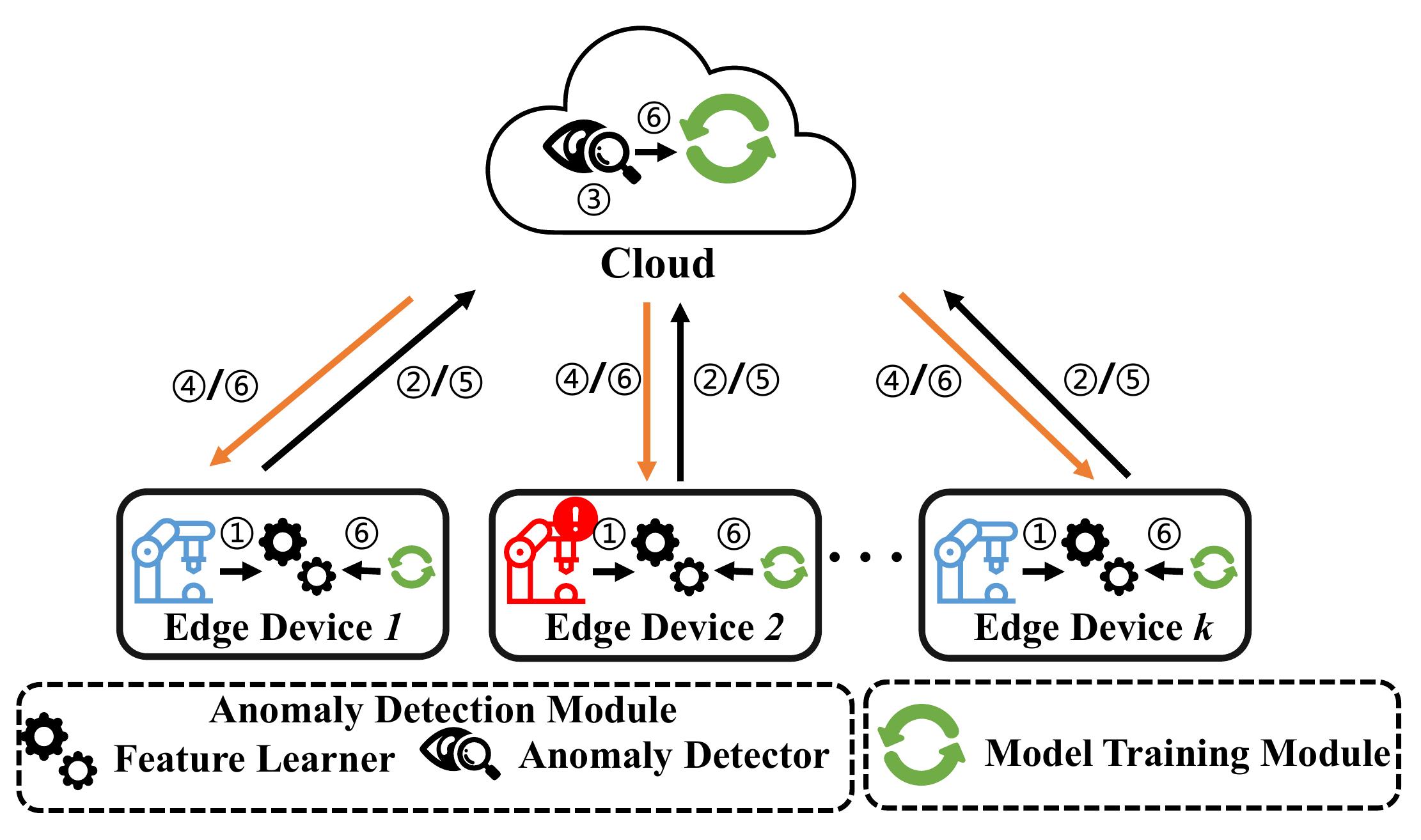}
	\caption{Overview of the proposed system model. The designed system model includes anomaly detection module and model training module. This system requires the following six operating steps (i.e., \textcircled{\scriptsize 1}--\textcircled{\scriptsize 6}): \textcircled{\scriptsize 1}The edge device uses the local feature learner to extract the features of the locally generated data; \textcircled{\scriptsize 2}The edge device uploads the extracted features to the cloud server; \textcircled{\scriptsize 3}The cloud server uses the cloud-side anomaly detector to evaluate the anomaly scores of these uploaded features; \textcircled{\scriptsize 4}The cloud server sends the anomaly scores to the edge devices; \textcircled{\scriptsize 5}--\textcircled{\scriptsize 6}The interactive gradient update between the cloud server and the edge devices. }
	\label{general framework}
\end{figure}

\section{System Model}

\subsection{System Architecture}
In this paper, we consider an MEC architecture involving a cloud server $\mathcal{S}$ and $\mathcal{K}$ edge devices. As shown in Fig. \ref{general framework}, the edge device is responsible for extracting abnormal feature representations using a local encoder on the local data set and uploading these features with the Gaussian noise to the cloud server, and the cloud server is responsible for using the decoder to calculate anomaly scores of the features uploaded by the edge devices for anomaly detection. Because the cloud-side model is different from the edge device-side model, the traditional FL training protocol is no longer suitable for such an anomaly detection framework. Furthermore, limited by privacy, communication overhead requirements, and timely detection of abnormal conditions, the previous MEC-based anomaly detection framework is not suitable for this task. Therefore, we need to design a new framework that is able to deal with the above problems. To this end, we design an FL-based anomaly detection framework for cloud manufacturing. Specifically, the framework includes two modules, namely anomaly detection, and model training. The main function of the anomaly detection module is to detect abnormal edge devices or abnormal data generated by edge devices. The main function of the model training module is to train the feature representation learner on the edge device side and the anomaly detector on the cloud side. Notably, we propose a new FL training protocol tailored for anomaly detection tasks in cloud manufacturing. Next, we introduce in detail the specific functions of the two modules designed as follows.

\noindent \textbf{Anomaly Detection Module.}
First, in order to solve the difficulty of timely and efficient anomaly detection in MEC environment, we customize an anomaly detection module in the proposed framework. In this module, the edge device firstly learns normal and abnormal feature representations through the local encoder, secondly, the edge device adds the Gaussian noise into the above features to protect data privacy and the edge device finally uploads the learned features to the cloud server instead of model updates. This means that our framework's training protocol is different from traditional FL training protocols. Then the cloud server uses the cloud-side decoder to decode the feature uploaded by the edge device and give anomaly scores. In this case, if the anomaly score exceeds the threshold $\tau$, the cloud server can promptly notify the customer or inform the maintenance staff to repair the machine. Indeed, our framework uses an encoder-decoder structure responsible for extracting and representing abnormal features. Encoder-decoder is very useful, simple, and effective feature extraction and learning structure. There are many variants of such a structure, such as auto-encoder and Transformer. Currently, many researchers utilize this structure for anomaly detection \cite{8478396,7539352}. Motivated by the above facts, the proposed anomaly detection framework applies this structure to anomaly detection. In addition, such an encoder-decoder structure has the advantage of controlling the output feature dimension, that is, it can reduce communication overhead on the premise of ensuring the performance of the model.

\noindent \textbf{Model Training Module.} 
In this module, we mainly introduce how to update the encoder-decoder structure of the framework. To fill the gap between the encoder-decoder structure training method and the FL training protocol, we improve a new FL training protocol to realize the efficient cooperation between the encoder-decoder structure (anomaly detection function) and FL. Specifically, when the cloud server uses the decoder to score the abnormality of the uploaded feature, the cloud server sends the corresponding anomaly score to the corresponding edge device. Then the edge devices use the label and anomaly score of the local data to calculate the losses and upload the losses to the cloud server. The cloud server uses an aggregation algorithm, \emph{i.e.}, FedAvg, to obtain the global loss and sends the loss to the edge devices. Finally, the edge devices and the cloud server update the encoder-decoder structure according to the loss and do not end the update until the loss converges.


\subsection{Discussion}
Here, we emphasize the difference between the proposed anomaly detection framework and the traditional MEC-based anomaly detection framework. Firstly, the proposed framework detects anomalies by scoring the uploaded features with the Gaussian noise instead of the original data, strengthening the framework's privacy protection capabilities and reducing communication overhead in the training process. Secondly, we use the decoder structure on the cloud and the encoder structure on the edge device-side for anomaly detection instead of using a model with the same structure. Specifically, we use an encoder-decoder structure with heterogeneous model characteristics (that is, the cloud-side model and the edge device-side model have different network structures) to detect anomalies. This means that the edge device and cloud server cannot update and train the global model through the aggregation of local model updates like the previous MEC framework (or FL framework). Training and updating heterogeneous models in FL is a challenging task. To this end, we design a new FL training protocol to overcome this model heterogeneity.

\begin{figure*}[t]
	\centering
	\subfigure [Anomaly Detection Module]{\includegraphics[width=0.49\linewidth]{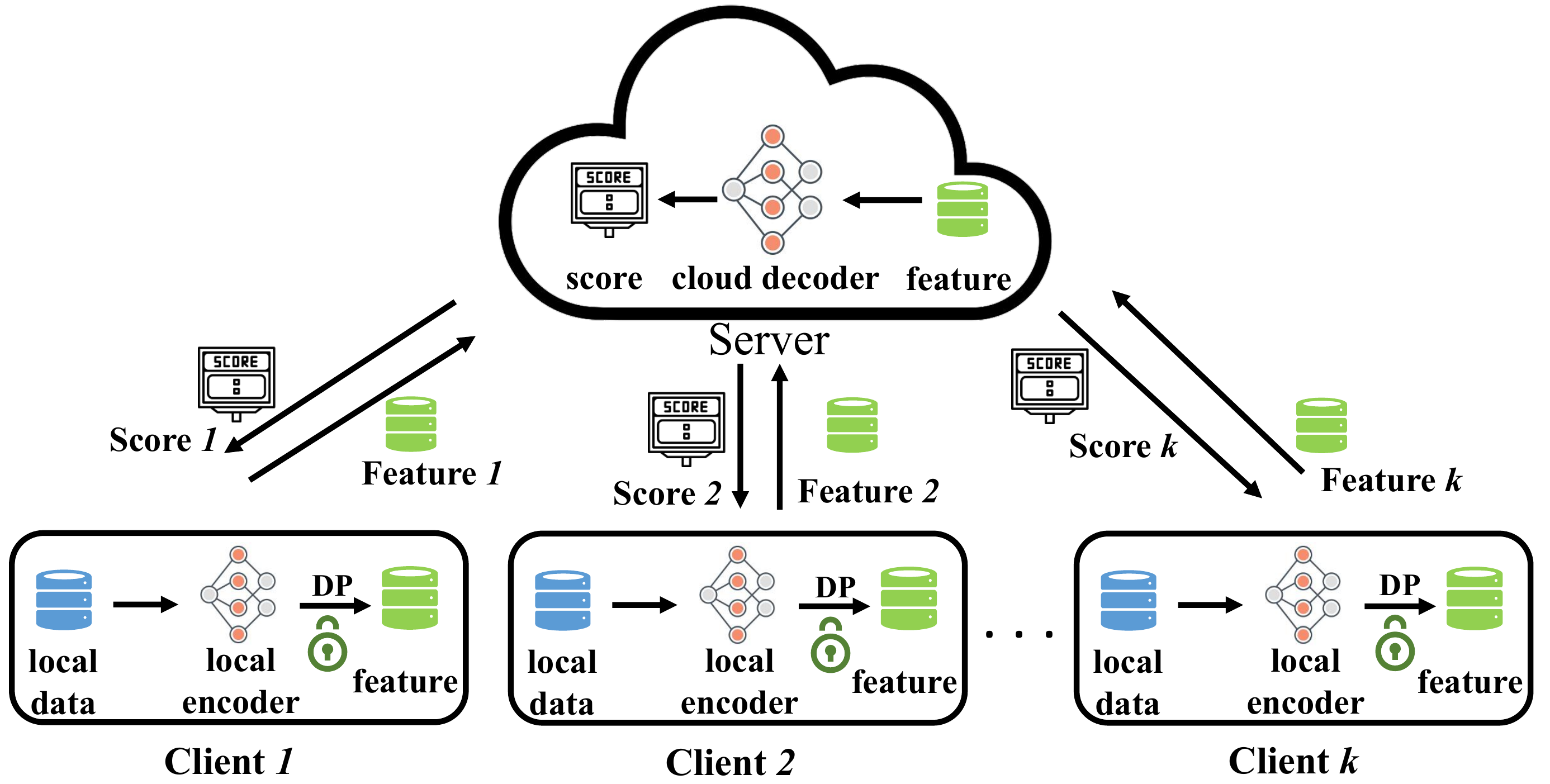}
		\label{anomaly detection step}}
	\subfigure[Model Training Module ]{\includegraphics[width=0.49
	\linewidth]{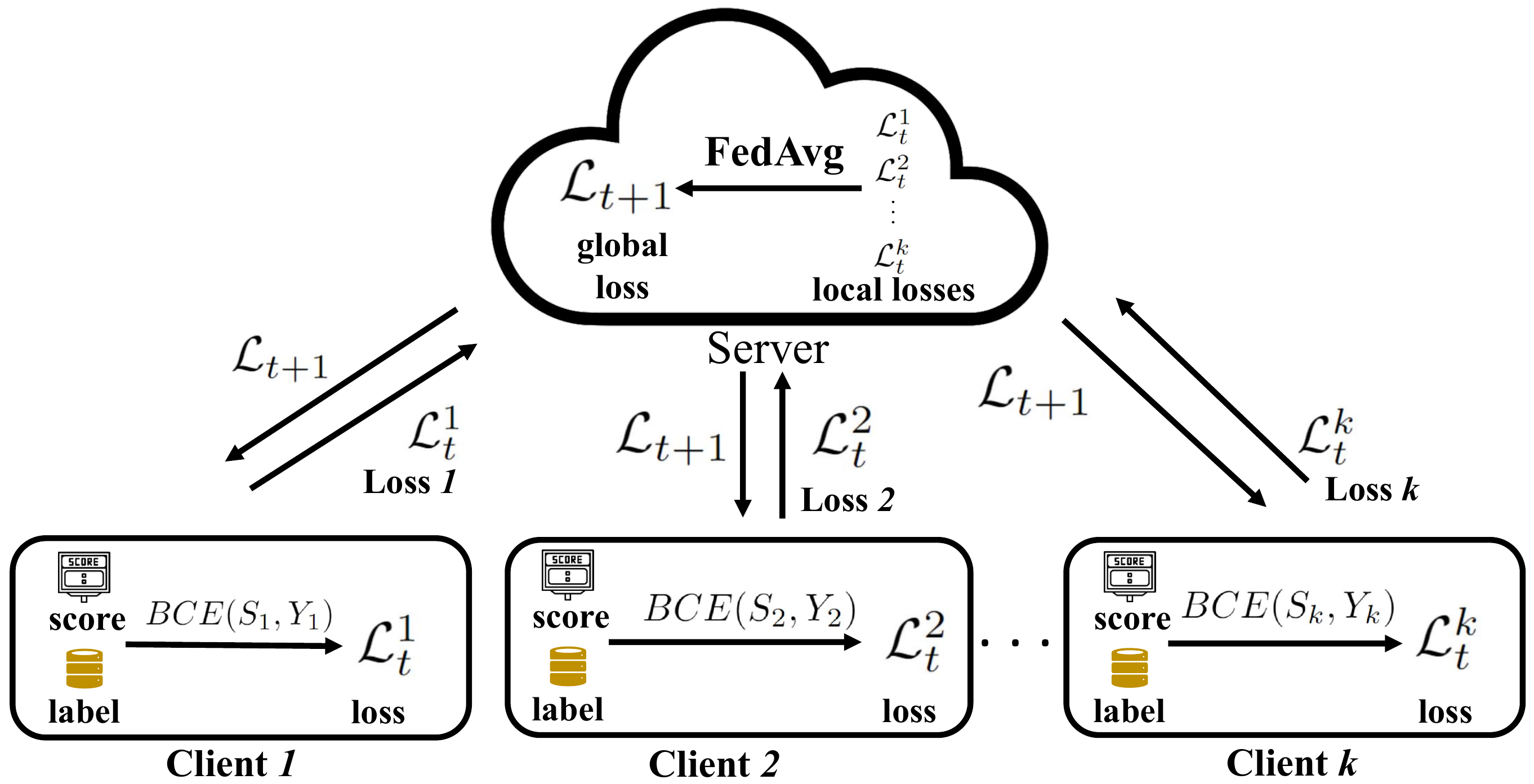}
		\label{Model Training step}}
	\caption{The framework consists of two entities: edge devices and cloud server. Specifically, each edge device holds the local data (in blue) and the corresponding label (in yellow), the features embedded in the local encoder (in green), and the local encoder. Furthermore, the cloud server holds a cloud decoder to perform anomaly score scoring. Finally, the edge devices and the server update their model by the global loss.}
	\label{FedAnomaly}
\end{figure*}

\section{Privacy-preserving Anomaly Detection in Cloud Manufacturing via Federated Transformer}
This section presents our Federated Transformer (\emph{i.e.}, FedAnomaly) framework design in detail. First, we give the problem definition of anomaly detection in cloud manufacturing. Second, we introduce the components of our proposed FedAnomaly framework. Finally, we provide the entire training process of the proposed framework.


\subsection{Problem Definition}
In the cloud manufacturing scenario, we consider a cloud server $\mathcal{S}$ to monitor the running status of $K$ edge devices (\emph{i.e.}, edge machines). In this context, once cloud server $\mathcal{S}$ detects that a specific edge device is faulty or generates abnormal data, it triggers a system alarm to notify the maintenance personnel to perform machine maintenance. We assume that the cloud server has a decoder $\phi^d$ model and each edge device has its own local data set $\mathcal{D}_k$ which has $n_k$ samples $\{(x_1,y_1), (x_2,y_2), \ldots, (x_{n_k},y_{n_k})\} \in \mathbb{R}^d$ and the local encoder $\phi_k^e$. The $k$-th edge device employs the $\phi_k^e$ to extract the anomaly feature representations can be expressed as follows:
\begin{equation}
    \mathcal{H}_k^i = \phi_k^e(x_i), x_i \in \mathcal{D}_k,
\end{equation}
where $\mathcal{D}_k$ is local data of the $k$-th edge device, $\phi_k^e$ is local encoder of the $k$-th edge device, and $\mathcal{H}_k$ has $n_k$ features $\{h_1, h_2, \ldots, h_{n_k}\} \in \mathbb{R}^m$ ($m<n$). 
Then, the $k$-th edge device apply DP technique to add Gaussian noise, \emph{i.e.}, as shown in Eq.\eqref{eq:differential_privacy_noise}, and the $k$-th edge device uploads the features $\mathcal{H}_k$ with Gaussian noise to the server. Given the features uploaded by edge devices \{$\mathcal{H}_1, \mathcal{H}_2, \ldots, \mathcal{H}_K$\} which the cloud server receives and employs $\phi^e$ to generate anomaly scores \{$\mathcal{S}_1$, $\mathcal{S}_2$ ,\ldots, $\mathcal{S}_K$\} for these features, it can be expressed as:
\begin{equation}
    \mathcal{S}_k^i = \phi^d(\mathcal{H}_k^i),
\end{equation}
where $\phi^d$ is the cloud decoder model, $\mathcal{S}_k$ is a set of anomaly scores $\{s_1, s_2, \ldots, s_{n_k}\}$ for $\mathcal{H}_k$, where a higher score shows that is more likely an anomaly and a lower score shows that is more likely a normal data.
Then, we pseudo-label the data through the threshold $\tau$, it can be expressed as:
\begin{equation}
\begin{cases}
    \hat{\mathcal{Y}_k^i} = 0,\text{ if } \mathcal{S}_k^i < \tau \\
    \hat{\mathcal{Y}_k^i} = 1,\text{ if } \mathcal{S}_k^i > \tau,
\end{cases}
\end{equation}
where $\hat{\mathcal{Y}_k}$ has $n_k$ pseudo-labels $\{\hat{y_1}, \hat{y_2}, \ldots, \hat{y_{n_k}}\}$. Then, the cloud server sends \{$\hat{\mathcal{Y}_1}$, $\hat{\mathcal{Y}_2}$ ,\ldots, $\hat{\mathcal{Y}_{K}}$\} to the edge device.
\par In this context, the objective function that we need to optimize is defined as follows:
\begin{equation}\label{eq-6}
\min L = \sum\limits_{k = 1}^K {\sum\limits_{i = 1}^{{n_k}} {l_k} } (\hat{\mathcal{Y}_k^i},\mathcal{Y}_k^i),
\end{equation}
where $l_k$ is the local objective function of $k$-th edge device, $K$ is the number of edge devices, $n_k$ is the number of samples in $\mathcal{D}_k$ of the $k$-th edge device.
\par The goal of this paper is accurately and timely detect anomalies in edge devices under the premise of protecting data privacy. To achieve the goal in our problem definition is to minimize Eq. \eqref{eq-6} in an efficient and timely manner. To this end, we need a method that can efficiently extract the coding features of anomalous data and perform accurate anomaly detection in a distributed fashion. In other words, we need to design an efficient feature extractor and propose an efficient collaborative training protocol to obtain ``better'' loss, thereby minimizing Eq. \eqref{eq-6} timely. Therefore, the insight of this paper is to combine the FL framework and Transformer to propose a novel collaborative training protocol to achieve the above goals.


\subsection{Model Design} \label{section: model}
In this section, we introduce our model design in the proposed FedAnomaly framework. Specifically, the proposed model architecture includes two parts: the feature representation learner on the edge device side, \emph{i.e.,} Transformer, and the anomaly scorer on the cloud server-side, \emph{i.e.,} MLP. The feature representation learner embeds the original data into the new feature space to obtain features that are conducive to distinguishing normal and abnormal data. The anomaly scorer learns to discriminate whether a given feature is normal or not. Next, we detail the functions of the above two parts.

\subsubsection{Feature Representation Learner}
The purpose of the feature representation learner is to generate the discriminative information that helps to separate anomalies from normal instances. In this paper, we use the Transformer as our feature representation learner because it utilizes an attention mechanism to embed the features of the data whlie better considering multiple attributes. In addition, to reduce communication overhead when uploading the features to the server, we compress the dimension of features by a fully connected layer. Therefore, the feature representation learner includes two parts: Transformer block (like the right of Fig.\ref{fig:transformer}) and compression block (the fully connected layer). The process of feature representation learner is as follows: firstly, the Transformer block embeds the original data into features through a self-attention mechanism, then the compression block reduces the dimension of features. It can be expressed as follows:

\begin{equation}
    \mathcal{H}_k=f_{FC}(\mathrm{Transformer\_Block}(D_k)),
\end{equation}
where $f_{FC}$ is the fully connected layer, Transformer\_Block is the attention information generated by the Transformer block, and $\mathcal{H}_k$ is a set of features derived from local data set of the $k$-th edge device.

\subsubsection{Differential Privacy Noise}
Although the feature representation learner embeds the original data from the original data space into a new feature space to change the original data distribution, it still has the security risk of leaking privacy. Therefore, we should adopt a more effective security mechanism to protect data privacy, that is, DP. The client should add the differential privacy noise into the features before uploading the features to the server. In this paper, we add the Gaussian noise into the features to ensure data privacy. According to the Eq.\eqref{eq:differential privacy} and Eq.\eqref{eq:differential_privacy_noise}, the security features are generated by:
\begin{equation}
    \mathcal{H}_k \leftarrow f(\mathcal{H}_k) + \mathcal{N}(0, \sigma^2 \Delta f^2).
\end{equation}
In addition, as a random variable, the feature approximately obeys Gaussian distribution, it means that $f(x)$ in Eq.\eqref{eq:differential_privacy_noise} is:
\begin{equation}
    f(x) = \frac{1}{\sqrt{2\pi\sigma}} \exp(-\frac{(x-\mu)^2}{2\sigma^2}) \cdot \epsilon.
\end{equation}
Furthermore, in order to satisfy the define of DP, $\sigma$ is subject to the following condition:
\begin{equation}
    \sigma =  \frac{\Delta f \cdot \sqrt{2 \ln(1.25/\delta)}}{\epsilon},
\end{equation}
where $\Delta f$ is the sensitivity of $f$ in Eq.\eqref{eq:differential_privacy_noise}. The features have stronger security after adding Gaussian noise and the edge device can upload the features to the server.

\subsubsection{Anomaly Scorer}
The purpose of the detector is to discriminate the features generated by the feature representation learner is normal or not. In this paper, we adopt MLP to score anomalies. The anomaly score is computed by:
\begin{equation}
    S_i=\mathrm{Sigmoid}(f_{MLP}(\mathcal{H}_i)),
\end{equation}
where $f_{MLP}$ is multi-layer perceptron at the cloud server, $\mathcal{H}_i$ is the features generated by the feature representation of the $i$-th edge device, and $S_i$ is anomaly score and the cloud server sends it to the edge device. In particular, the higher anomaly scores represent a higher likelihood that data are anomaly data and the lower anomaly scores represent a lower likelihood that data are normal.

\subsubsection{Loss Function}
There have two parts in FedAnomaly, the feature representation learner and the anomaly scorer. They are optimized through the same objective function during the training process. In this paper, we employ a binary cross-entropy loss function to train both. We set the label of anomalies to 1, \emph{i.e.}, $y_{a}=1$, and set the label of normal data to 0, \emph{i.e.}, $y_n=0$. The purpose of this is to allow the anomaly scorer to discriminative score that the score of abnormal data is close to 1 and the score of normal data close to 0. The loss function can be represented as follows:
\begin{equation}
    L=-\sum_{i=1}^n y_i log(x_i) + (1-y_i) log(1-x_i),
\end{equation}
where $y_i$ is the label of the $i$-th sample and $x_i$ is the anomaly score of the $i$-th sample. The edge devices and cloud server calculate the loss based on the anomaly scores and use this loss to update the model. We present our new training protocol in Algorithm. \ref{algorithm} as follows:

\begin{algorithm}[!t]
	\caption{FedAnomaly Training Protocol}\label{algorithm}
	\begin{algorithmic}[1]
	\REQUIRE $K$ is the total number of edge devices, $D_k$ is the local data set of the $k$-th edge device, and $\eta$ is the learning rate.
	\ENSURE The optimal the local encoder $\phi^e$ and the cloud decoder $\phi^d$.
	\STATE Initialize $\phi^e$ and $\phi^d$
	\FOR{each round $t = 0, 1, \ldots, n$}
	   \STATE $H$ = \textbf{Edge\_Feature\_Learning}
	   \IF{Add Gaussian Noise}
	        \STATE $H = f(H)+ \mathcal{N}(0, \sigma^2 \Delta f^2)$
	   \ENDIF
	   \STATE $S$ = \textbf{Cloud\_Anomaly\_Scoring($H$)}
	   \STATE \textbf{Model\_Training($S$)}
	\ENDFOR
	\STATE \textbf{Edge\_Feature\_Learning:}
    	\FOR{each edge device $k = 0, 1, \ldots, K$}
    	    \STATE $H_k = \phi^e_{k,t}(D_k)$
    	\ENDFOR
    	\RETURN $H$ to cloud server. $//$ $H$ = \{$H_1, H_2, \ldots, H_K$\}
    \STATE \textbf{Cloud\_Anomaly\_Scoring($H$):}
    \FOR{each batch $k = 0, 1, \ldots, K$}
        \STATE $S_k = \phi^d_t(H_k)$
    \ENDFOR
    \RETURN $S$ to the edge device. $//$ $S$ = \{$S_1, S_2, \ldots, S_K$\}
    \STATE \textbf{Model\_Training($S$):}
    \FOR{each client $k=0,1,2, \ldots, K$}
        \STATE ${L_{k}}=BCE(Y_k,S_k)$ $//$ $Y_k$ is the label of $D_k$, client sends $L_{k}$ to the server
    \ENDFOR
    \STATE $L=\sum_{k=1}^K \frac{1}{n_k} L_{k}$ $//$ Run on the server
    \STATE $\phi_{t+1}^{d} = \phi_{t}^{d} - \eta \bigtriangledown L_k(\phi_{t}^d)$ $//$ Update the cloud decoder
    \FOR{each edge device $k = 0, 1, \ldots, K$}
        \STATE $\phi_{k,t+1}^{e} = \phi_{k,t}^{e} - \eta \bigtriangledown L_k(\phi_{k,t}^e)$ $//$ Update the local encoder
    \ENDFOR
	\end{algorithmic}
\end{algorithm}

\begin{itemize}
    \item \textbf{\textit{Step 0, Initialization:}} The cloud server and edge device initialize their models separately.
    \item \textbf{\textit{Step 1, Anomaly Detection:}} The edge devices firstly use the local encoder model to embed new feature representations $\mathcal{H}_k$ on the local data set $D_k$, then the edge devices add Gaussian noise into the features and finally upload $\mathcal{H}_k$ to the cloud server. Then it uses the cloud decoder to generate anomaly score $\mathcal{S}_k$ for the uploaded features to complete the anomaly detection task and send $\mathcal{S}_k$ to the edge device.
    \item \textbf{\textit{Step 2, Model Training:}} The edge device computes the local loss $\mathcal{L}_t^k$ through the anomaly scores $\mathcal{S}_k$ and labels of local data set $D_k$, and uploads $\mathcal{L}_t^k$ to cloud server. The cloud server collects these losses and uses the FedAvg \cite{mcmahan2017communication} to obtain the global loss $\mathcal{L}_{t+1}$, \emph{i.e.}, ${\mathcal{L}_{t + 1}} = \sum\limits_{k = 1}^K {\frac{{|{D_k}|}}{{|\mathcal{D}|}}} \mathcal{L}_t^k$. Then the cloud server sends $\mathcal{L}_{t+1}$ to the selected edge devices. So the edge devices and cloud server update their model through $\mathcal{L}_{t+1}$.
\end{itemize}

\section{Experiment}

\subsection{Experiment Setup}
To evaluate the performance of the proposed model, we conduct extensive experiments on four representative public data sets. 

\noindent \textbf{Data set.} In this paper, we adopt four real-world data sets for evaluations, \emph{i.e.}, NSL-KDD \cite{5356528}, Spambase\footnote{\url{https://archive.ics.uci.edu/ml/data sets/spambase}}, Arrhythmia\footnote{\url{https://archive.ics.uci.edu/ml/data sets/arrhythmia}}, and Shuttle\footnote{\url{https://archive.ics.uci.edu/ml/data sets/Statlog+(Shuttle)}} data set. The data sets cover number of normal data, number of abnormal data, and dimensions, as shown in Table \ref{tab: data set}, allowing us to effectively explore the anomaly detecting utility of the proposed model. 


\begin{table}[H]
\caption{data set}
\label{tab: data set}
\centering
\begin{tabular}{cccc} 
\hline
Data set    & Normal data & Abnormal data & Dimension  \\ 
\hline
NSL-KDD    & 71,463    & 77,054        & 122        \\
Spambase   & 2,788     & 1,813         & 57         \\
Arrhythmia & 386       & 66            & 279        \\
Shuttle    & 45,586    & 3,511         & 9          \\
\hline
\end{tabular}
\end{table}

\noindent \textbf{Data Partition.}
In this experiment, we consider the distribution of anomalies in the real world. We randomly select the $80\%$ normal samples and 30 abnormal samples (because the number of samples in Arrhythmia is too small, we randomly select 15 samples in Arrhythmia) into the training data set (containing a small number of samples marked as anomalies), and divide the $20\%$ normal and abnormal samples into the test data set. Furthermore, due to the anomaly detection task is usually weakly-supervised learning, so we partition the data according to the setting of weakly-supervised learning. Firstly, we consider that there are unlabeled anomalies in the real world, we fix $2\%$ of the unlabeled anomalies in the training data set as noises. Secondly, referring to the settings of baseline experimental data set, the training batch has half of the unlabeled samples and half of labeled anomalies samples. Thirdly, due to the unlabeled abnormal samples being uncommon in the real world, we set all unlabeled data as the normal data.


\noindent \textbf{Baselines.}
We choose the SOTA weakly-supervised anomaly detection method \cite{9465358}, \emph{i.e.}, Feature Encoding with AutoEncoders (FEAE), and the centralized implementation of our method, \emph{i.e.}, Transformer for Anomaly Detection (TAD), as our baselines. Firstly, in \cite{9465358}, the authors designed a weakly-supervised anomaly detection framework (FEAE) which contains two modules: feature encoder module and anomaly score generator module. Specifically, \cite{9465358} uses the pre-trained autoencoder as the feature encoder, and \cite{9465358} utilizes the feature encoder to obtain features that are hidden representation, reconstruction error, and the residual reconstruction vector. Especially, the hidden representation is obtained by the encoder of autoencoder encodes the original data from initial space to the new feature space, reconstruction error is the Euclidean norm obtained by the original data minuses the output by the decoder and the reconstruction residual vector is obtained by the original data minuses the output by decoder divided by the reconstruction error. Then the anomalies are distinguished through anomaly scores obtained by the anomaly score generator through the above features. Secondly, TAD is the centralized implementation of our method. Because we want to explore the difference between the performance of our model in centralized and distributed, so we set TAD as our baseline.

\noindent \textbf{Parameters.}
Both the proposed model and baselines use Adam to optimize the network with a fixed learning rate $\eta=10^{-4}$, and we set the batch size of the training data set, and testing data set $b = 32$. In FedAnomaly, we consider the consistency of the data sample when comparing with the baseline, \emph{i.e.}, unlabeled samples and labeled samples have the same amount of data, so we set the number of total edge devices $k=3$, and the participation rate is $1$. That is to say, and the data is divided into $3$ equal parts.

\noindent \textbf{Evaluation Metrics.}
In the real world, abnormal samples are very uncommon, it leads that is unreasonable to evaluate the performance of the model by accuracy, so we evaluate our model performance with two metrics, \emph{i.e.}, AUC-ROC, among them, the y-axis as $TPR=\frac{TP}{(TP+FN)}$ and x-axis as $FPR=\frac{FP}{(FP+TN)}$ and AUC-PR, among them, the y-axis as $P=\frac{TP}{(TP+FP)}$ and x-axis as $R=\frac{TP}{(TP+FN)}$. For these two metrics, the larger value means the better performance of the model.

\subsection{Evaluation}
In this section, we evaluate FedAnomaly by conducting an experimental comparative analysis, and then we perform ablative analysis, it allows us to understand the model performance in different settings.

\begin{table*}[t!]
\centering
\caption{Performance comparison of the four different data sets.}
\label{tab: performance}
\begin{tabular}{|c|ccc|ccc|} 
\hline
\multirow{2}{*}{Data set} & \multicolumn{3}{c|}{AUC-ROC} & \multicolumn{3}{c|}{AUC-PR} \\ 
\cline{2-7} & FedAnomaly & TAD & FEAE & FedAnomaly & TAD & FEAE \\ \hline
NSL-KDD & 0.976$\pm$0.012 & \textbf{0.979$\pm$0.011} & 0.959$\pm$0.014 & 0.975$\pm$0.006 & \textbf{0.978$\pm$0.010} & 0.970$\pm$0.003 \\
Spambase & 0.945$\pm$0.016 & \textbf{0.947$\pm$0.003} & 0.921$\pm$0.015 & 0.915$\pm$0.014 & \textbf{0.934$\pm$0.005} & 0.910$\pm$0.007  \\
Shuttle & \textbf{0.994$\pm$0.005} & 0.993$\pm$0.004 & 0.983$\pm$0.011 & \textbf{0.976$\pm$0.008} & \textbf{0.976$\pm$0.012} & 0.975$\pm$0.006   \\
Arrhythmia & 0.907$\pm$0.135 & \textbf{0.946$\pm$0.015} & 0.868$\pm$0.195 & 0.786$\pm$0.126 & \textbf{0.797$\pm$0.015} & 0.724 $\pm$0.017  \\ \hline
Average & 0.955$\pm$0.042 & \textbf{0.966$\pm$0.0085} & 0.954$\pm$0.0587 & 0.913$\pm$0.0385 & \textbf{0.921$\pm$0.0105} & 0.894$\pm$0.0082 \\\hline
\end{tabular}
\vspace{-0.5cm}
\end{table*}

\subsubsection{Performance Analysis} 
\begin{figure}[!t]
	\centering
	\large
	\subfigure [AUC-ROC]{\includegraphics[width=0.46\linewidth]{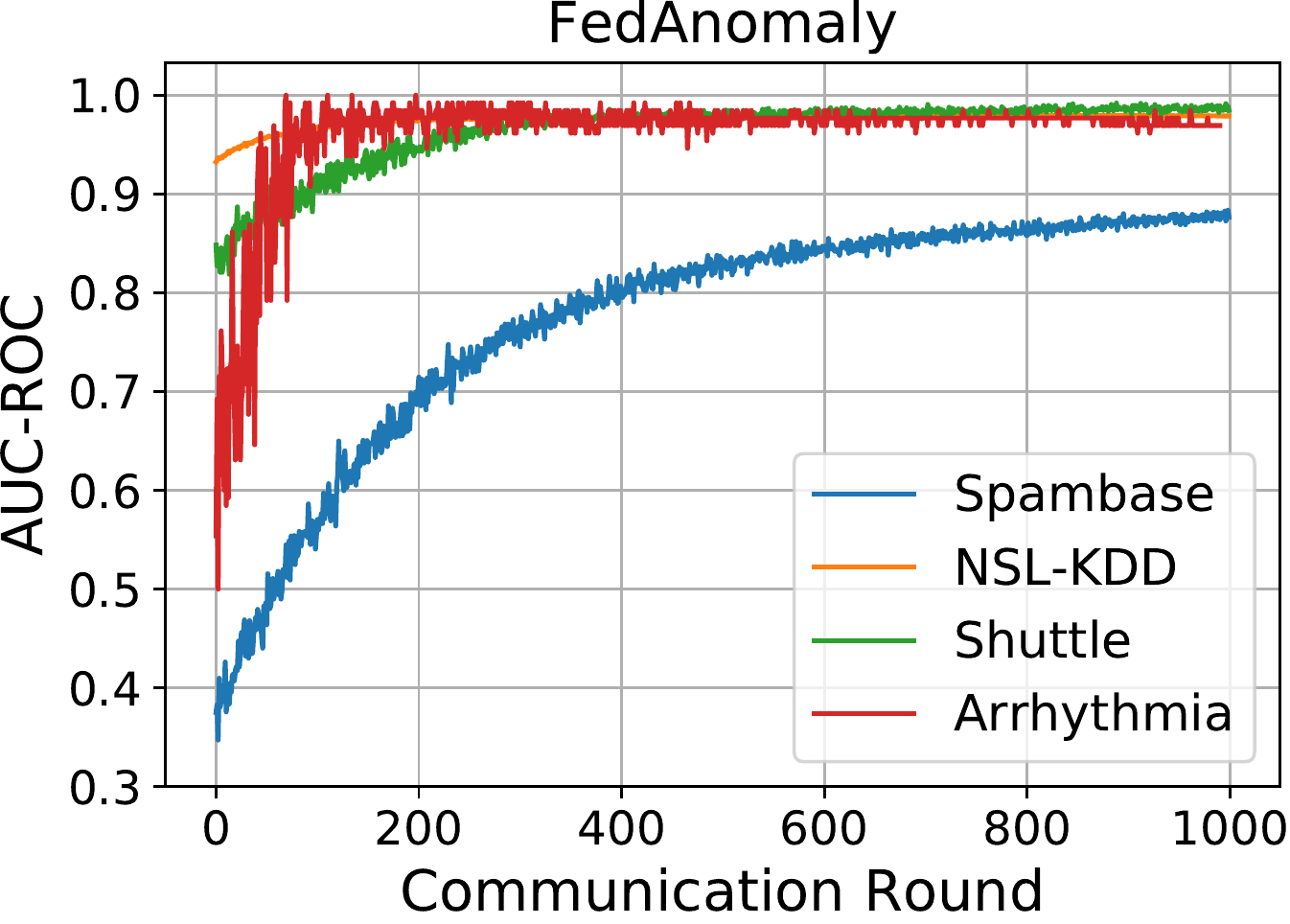}
		\label{2-1}}
	\hfill
	\subfigure[AUC-PR]{	\includegraphics[width=0.46\linewidth]{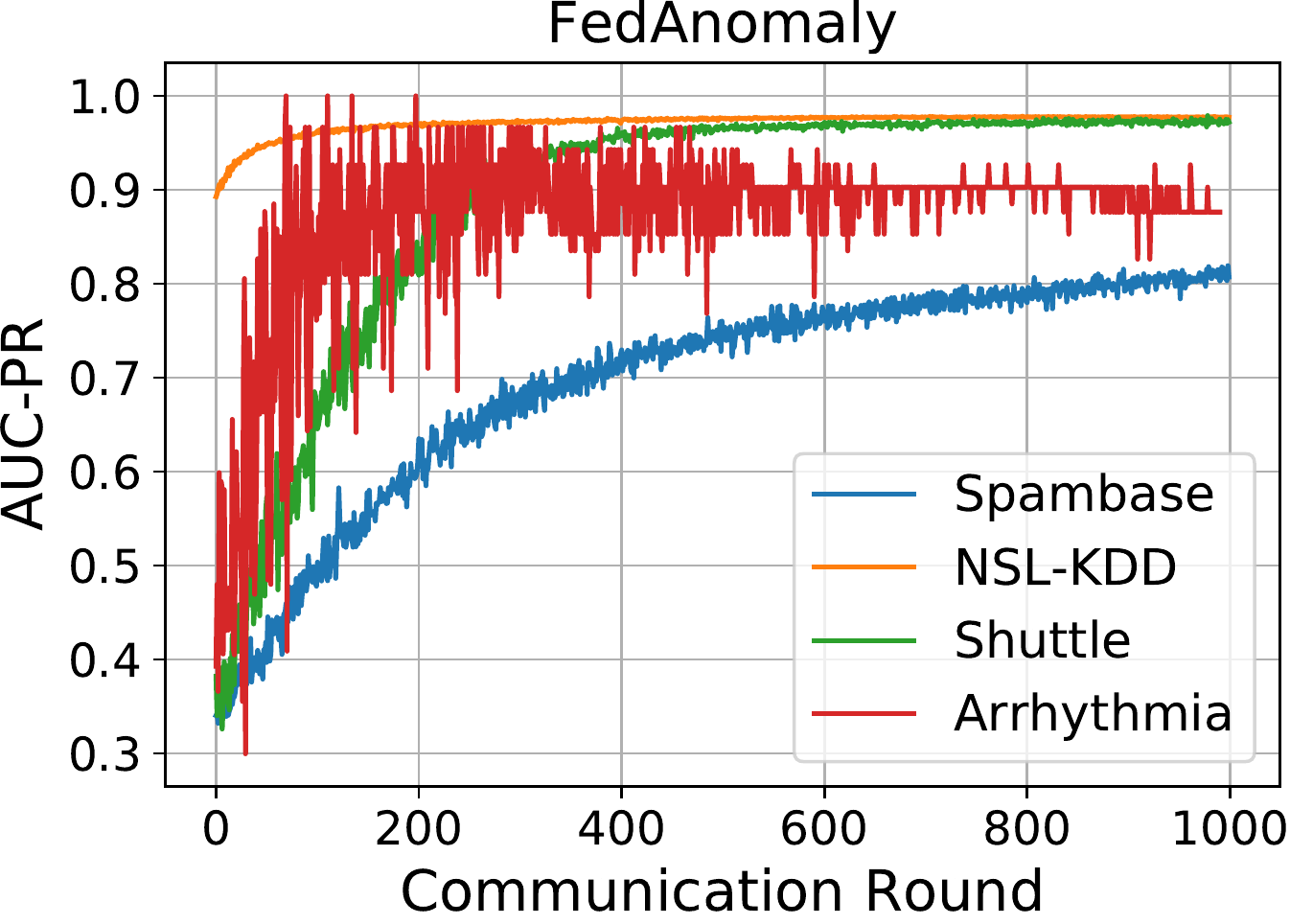}
		\label{2-2}}
	\caption{Performance comparison of different data sets by using two metrics.}
	\label{fig: ex1}
\end{figure}

We evaluate the proposed model on four benchmark data sets. Table \ref{tab: performance} reports the performance of the model after convergence, and the best results in each data set are highlighted in bold. Next, we compare FedAnomaly with the baselines. From the experimental results show that FedAnomaly is better than FEAE on the four data sets. Specifically, for AUC-ROC and AUC-PR metrics, the proposed method is better than FEAE by about $1\%$. In addition, the performance of FedAnomaly on the NSL-KDD, Spambase, and Shuttle data sets is slightly lower than that of TAD. This is because the data of FedAnomaly may be non-IID, which leads to slight performance degradation. Fig. \ref{fig: ex1} shows the performance curve of FedAnomaly on four benchmark data sets. As shown in Fig. \ref{fig: ex1}, the convergence speed of FedAnomaly on the NSL-KDD, Shuttle, and Arrhythmia data sets is similar, \emph{i.e.}, the model performance tends to be stable after 400 communication rounds. These experimental results show that the FedAnomaly can reach convergence in about 300 communication rounds. Furthermore, we observe that the performance of FedAnomaly fluctuates slightly on the NSL-KDD, Spambase, and Shuttle data sets. However, due to the small number of abnormal samples in Arrhythmia, the model is prone to overfitting and large fluctuations. Overall, the above numerical analysis shows that FedAnomaly is stable in most cases.

\subsubsection{Effect of the Number of Edge Devices} In this experiment, we explore the effect of the number of edge devices, \emph{i.e.}, $C$, on the performance of FedAnomaly. To this end, we fix $C=\{3,6,10\}$ in our framework to conduct experiment. We combine two metrics, \emph{i.e.}, the performance of the model and communication rounds that model achieves convergence, to evaluate our framework's performance. As shown in the Table \ref{tab: client}, as the number of edge devices increases, the performance of FedAnomaly decreases slightly, but the performance loss is only within $1\%$. In addition, we find that as the number of edge devices increases, the number of communication rounds required for the model to reach convergence also increases. The reason is that the data distribution of each edge device is different, so the more edge devices, the more serious the non-IID level of the data, and the slower the model convergence speed. The above experimental results show that FedAnomaly can be applied to cloud manufacturing for anomaly detection due to its stable scalability.

\begin{table}
\centering
\caption{Performance comparison of the different edge devices numbers ($C$ means the number of edge devices).}
\label{tab: client}
\begin{tabular}{|c|c|c|c|c|} 
\hline
\multicolumn{2}{|c|}{\diagbox{$C$}{data set}} & NSL-KDD & Spambase & Shuttle  \\ 
\hline
\multirow{4}{*}{3}& AUC-ROC & 0.9765 & 0.9469 & 0.9937 \\ 
\cline{2-5} & Rounds & 388 & 4381 & 2610 \\ 
\cline{2-5} & AUC-PR & 0.9753 & 0.9167 & 0.9798 \\ 
\cline{2-5} & Rounds & 388 & 4907 & 2626 \\ \hline
\multirow{4}{*}{6} & AUC-ROC & 0.9766 & 0.9465 & 0.9939 \\ 
\cline{2-5} & Rounds & 1117 & 4792 & 2961 \\ 
\cline{2-5} & AUC-PR & 0.9752 & 0.9162 & 0.9781 \\ 
\cline{2-5} & Rounds & 1083 & 4799 & 2581 \\ \hline
\multirow{4}{*}{10} & AUC-ROC & 0.9762 & 0.9377 & 0.9912 \\ 
\cline{2-5} & Rounds & 2380 & 5126 & 4169 \\ 
\cline{2-5} & AUC-PR & 0.9750 & 0.9097 & 0.9781 \\ 
\cline{2-5} & Rounds & 2380 & 5126 & 3992 \\\hline
\end{tabular}
\end{table}

\subsubsection{Effect of the Features Dimensionss}
\begin{figure}[t]
	\centering
	\subfigure {\includegraphics[width=0.48\linewidth]{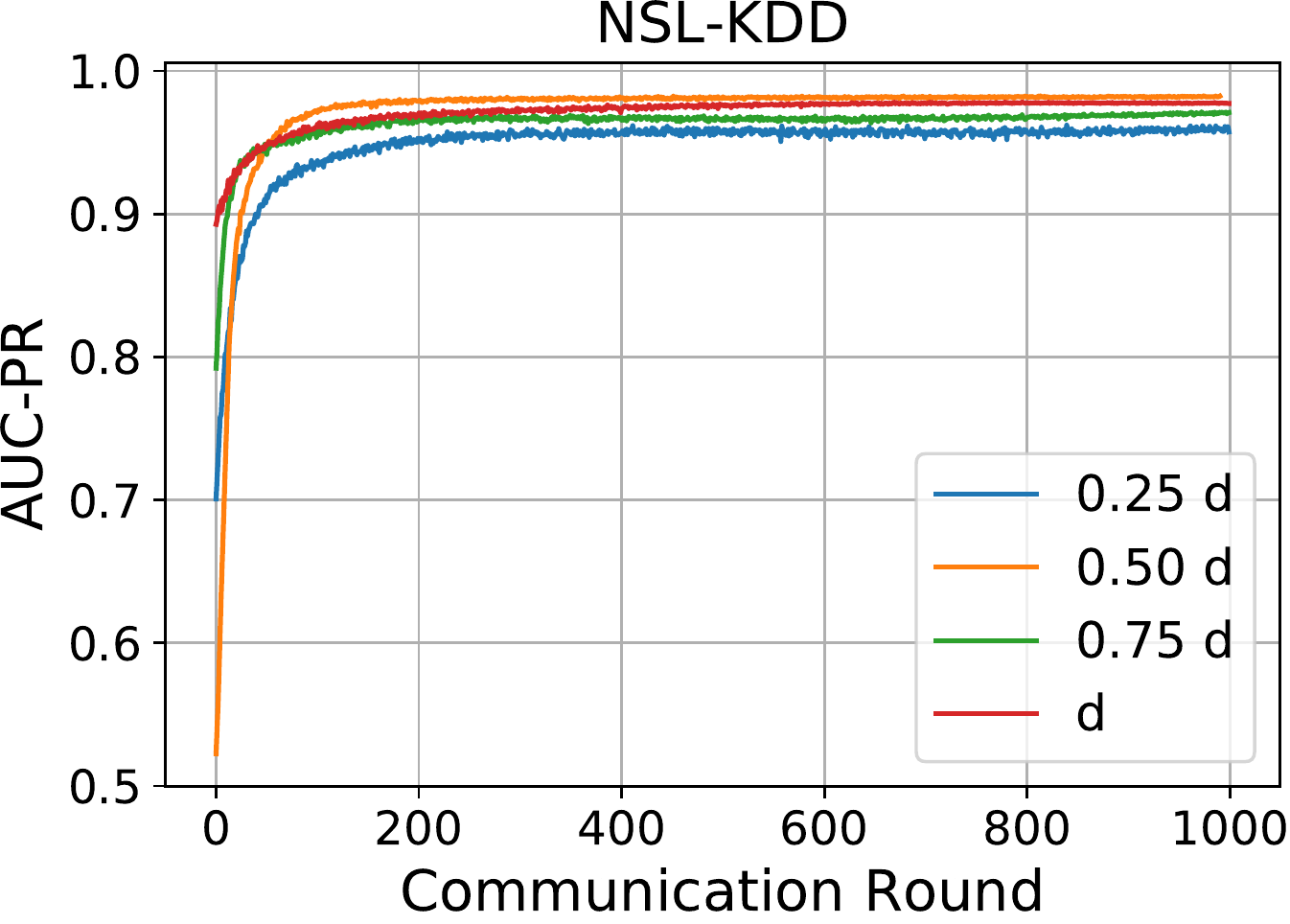}
		\label{3-1}}
	\subfigure{\includegraphics[width=0.48\linewidth]{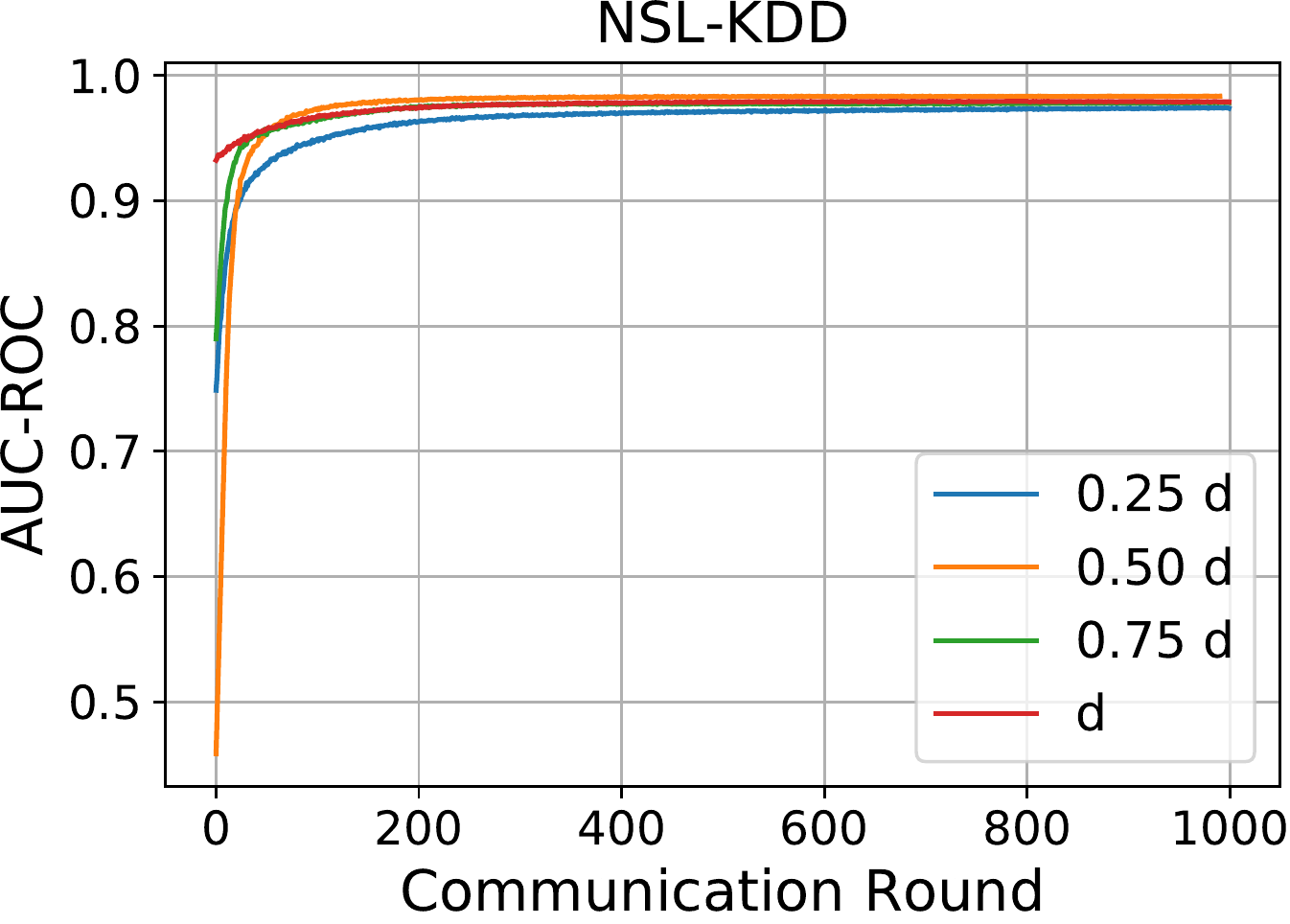}
		\label{3-2}}
	\subfigure{\includegraphics[width=0.48\linewidth]{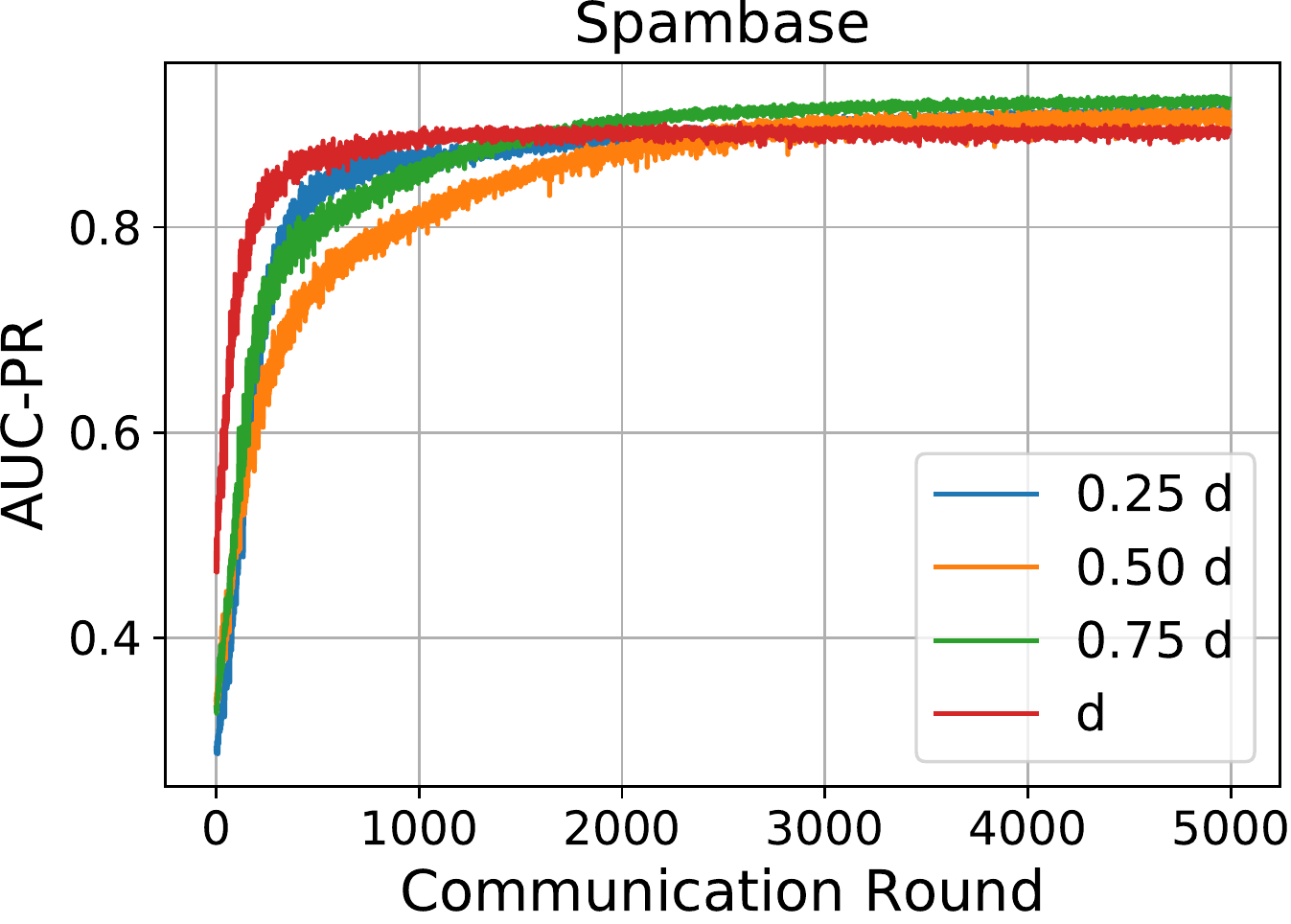}
		\label{3-3}}
	\subfigure{\includegraphics[width=0.48\linewidth]{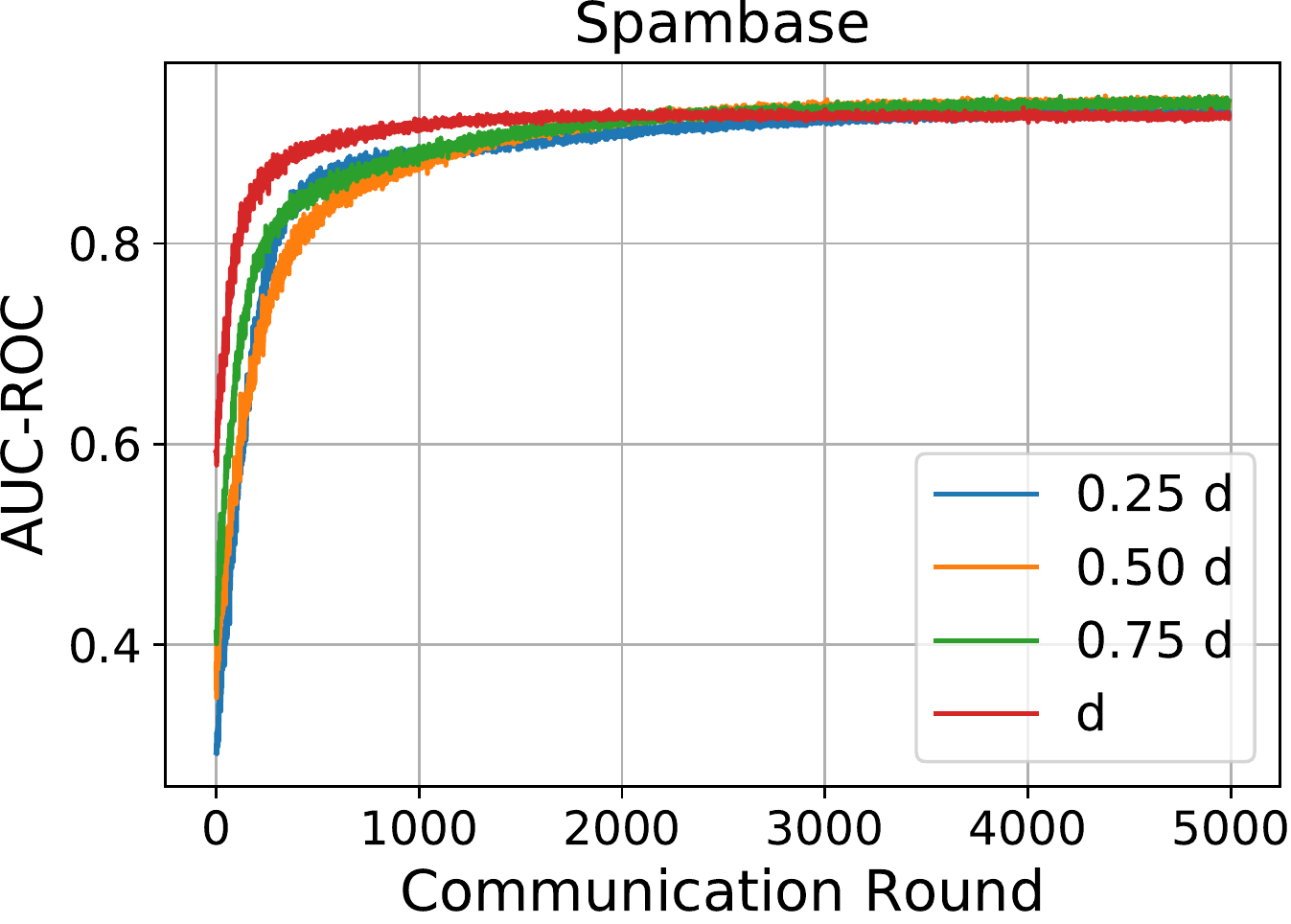}
		\label{3-4}}
	\subfigure{\includegraphics[width=0.48\linewidth]{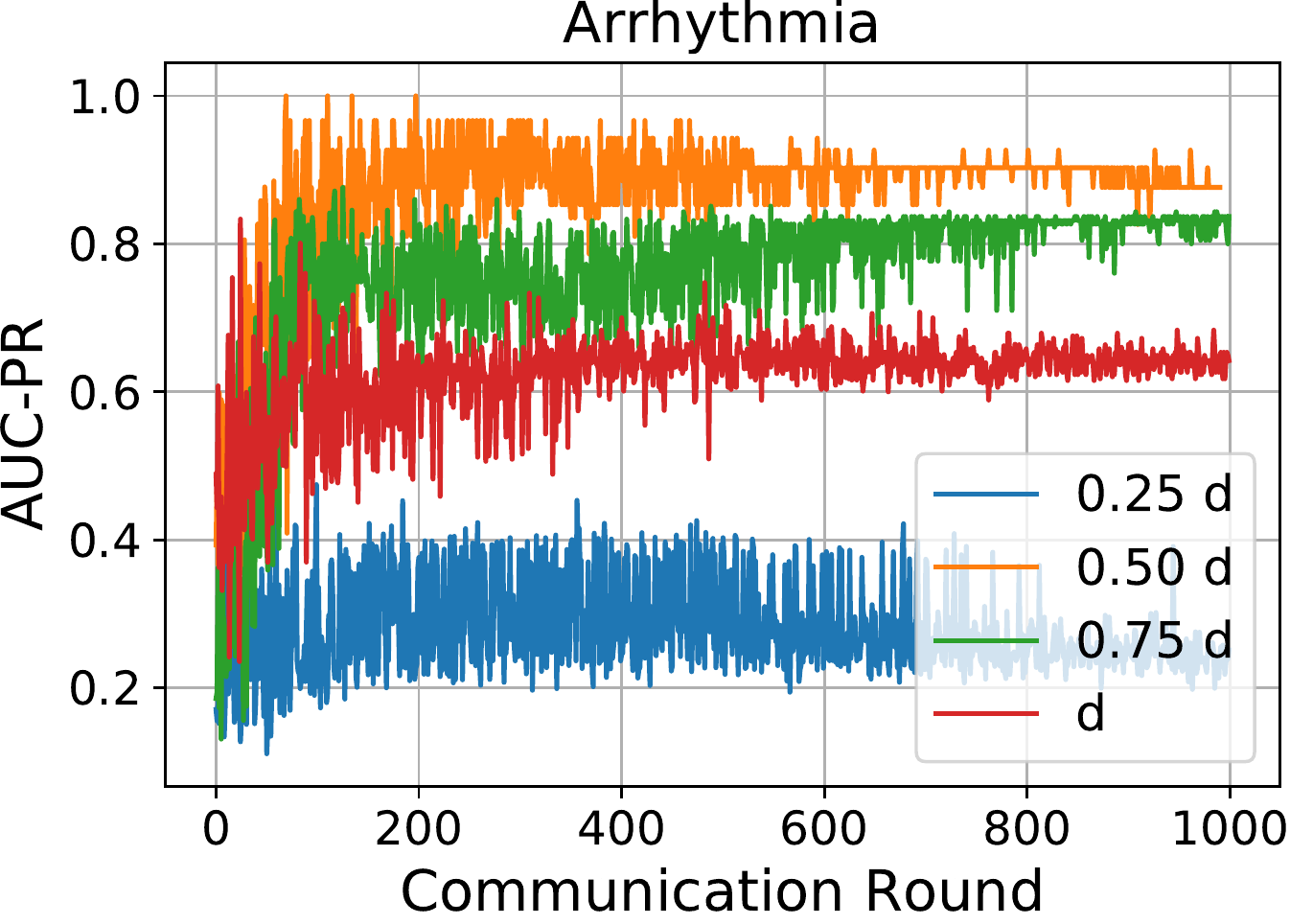}
		\label{3-5}}
	\subfigure{\includegraphics[width=0.48\linewidth]{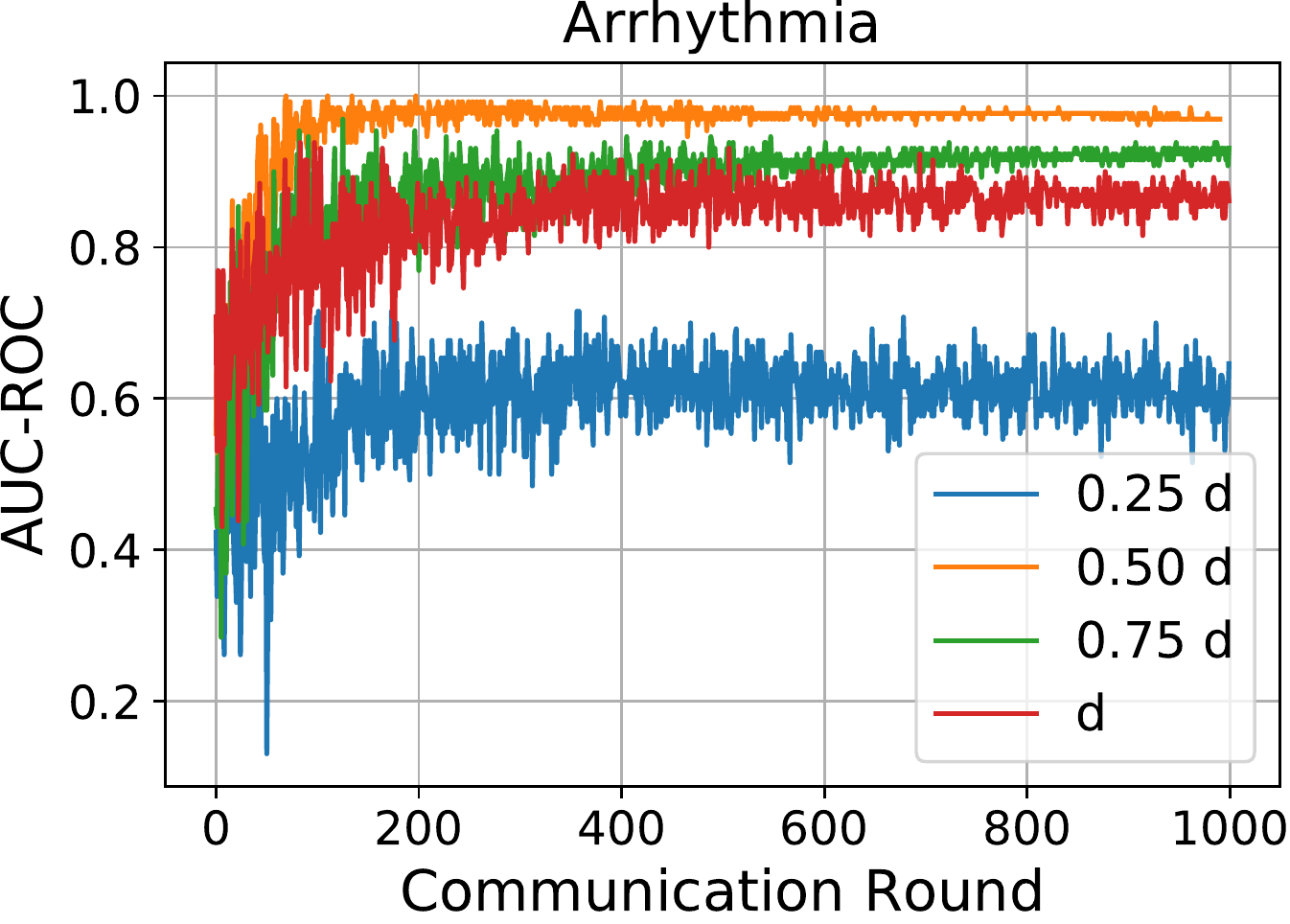}
		\label{3-6}}
	\caption{Performance comparison of the different dimension.}
	\label{dim}
\end{figure}

Recall that, in FedAnomaly, edge devices need to upload features to the cloud server for anomaly detection. Therefore, the dimensions of the features that need to be uploaded determine the system's communication overhead and performance. To this end, we need to study the impact of feature dimensions on model performance and communication overhead to explore the balance between model performance and communication overhead. Specifically, we set the dimension of the feature to be uploaded $d_f=\{0.25d, 0.5d, 0.75d, d\}$, where $d$ is the dimension of the original data. We evaluate the impact of feature dimension $d$ by observing the system's performance on four data sets and the number of communication rounds. As shown in Fig. \ref{dim}, we find that the communication rounds required for model convergence under different feature dimensions $d$ on the NSL-KDD data set are roughly the same. When $d_f=0.25d$ and $d$, in terms of AUC-PR, the performance of the model under different feature dimensions is close, but they are both better than the performance when $d_f=0.25d$ and $0.75d$. However, as far as AUC-ROC is concerned, when $d_f=0.75d$ and $d$, the performance of the model is similar, but their performance is better than that when $d_f=0.25d$ and $0.5d$. For the Spambase data set, when $d_f=0.5d$, the convergence speed of the model is slower than in other cases. In addition, when $d_f=0.5d$ and $0.75d$, in terms of AUC-PR and AUC-ROC, the model performance is similar and better than the model performance under other feature dimensions cases. For the Arrhythmia data set, the convergence speed is similar in all cases due to the small number of data sets. However, when $d_f=0.5d$ and $0.75d$, the model's performance is better than that in other cases. From the above experimental results, we believe that the best setting of the feature dimension in FedAnomaly is $d_f=0.5d$. 
\par
\begin{figure}[t]
	\centering
	\includegraphics[width=0.7\linewidth]{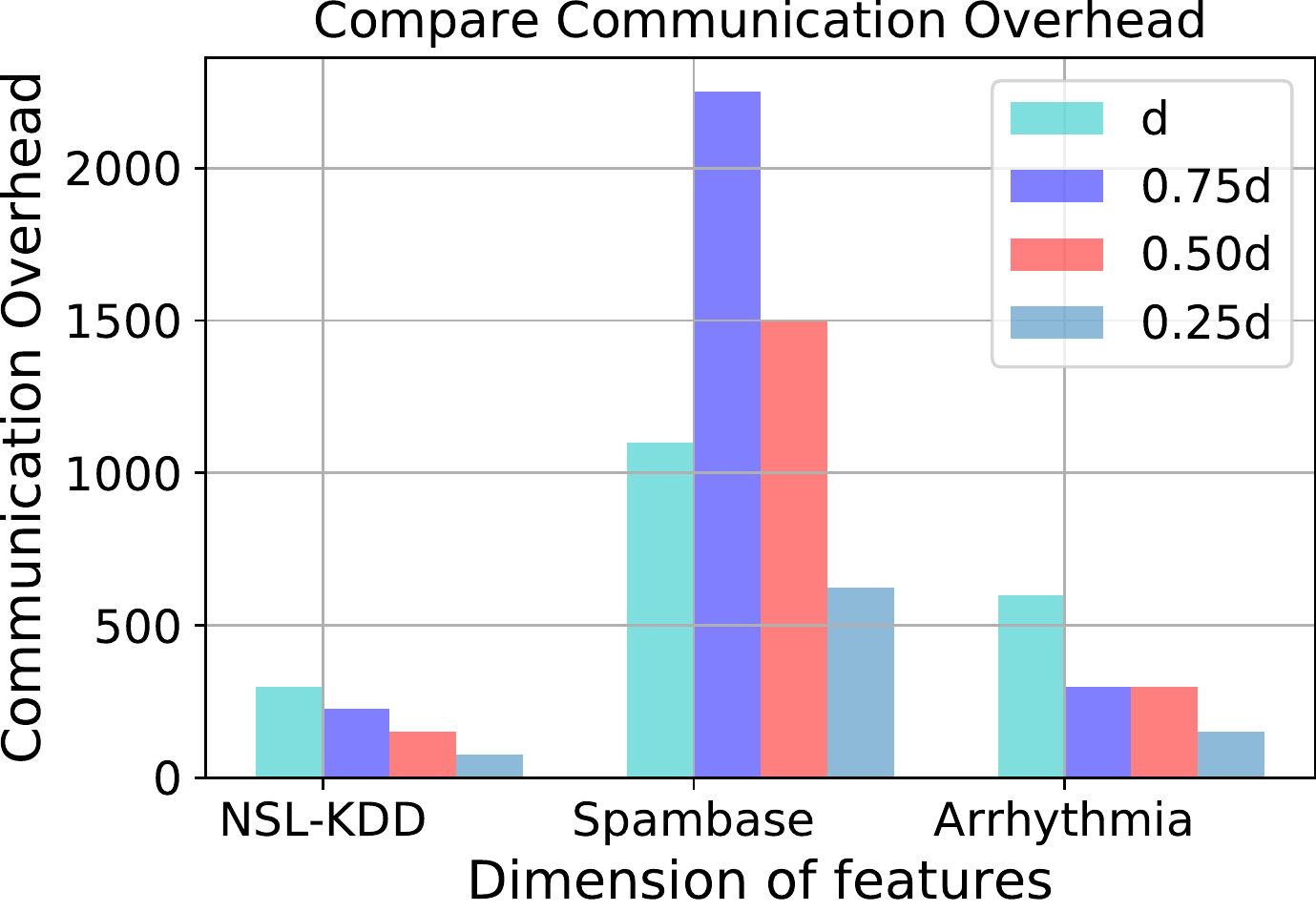}
	\caption{Communication overhead comparison of the different dimension}
	\label{fig: communication overhead}
\end{figure}

 In this setting, we compare the communication overhead in different dimension of features, as shown in Fig.\ref{fig: communication overhead}. The y-axis of the Fig.\ref{fig: communication overhead} means the communication overhead, we set the communication overhead of each round is $1$ when the dimension of features is $d$. We can observe that the framework has the least communication overhead between $d_f=0.25d$ and $d_f=0.5d$.

\begin{table}
\caption{Performance of Adding Gaussian Noise.}
\centering
\refstepcounter{table}
\label{tab: differential privacy}
\begin{tabular}{|c|c|c|c|c|} 
\hline
\multicolumn{1}{|l|}{\multirow{2}{*}{Data set}} & \multicolumn{2}{c|}{AUC-ROC} & \multicolumn{2}{c|}{AUC-PR}  \\ 
\cline{2-5}
\multicolumn{1}{|l|}{} & DP & FedAnomaly & DP & FedAnomaly           \\ 
\hline
NSL-KDD     & 0.974 & 0.976   & 0.976 & 0.975  \\
Spambase    & 0.931 & 0.945   & 0.909 & 0.915  \\
Shuttle     & 0.986 & 0.994   & 0.972 & 0.976  \\
Arrhythmia  & 0.885 & 0.907   & 0.766 & 0.786  \\
\hline
\end{tabular}
\end{table}

\subsubsection{Effect of Adding Gaussian Noise}
We add Gaussian noise into the features uploaded to protect data privacy. We evaluate the model performance after adding Gaussian noise, and we compare the performance of the original framework without Gaussian noise and the framework with Gaussian noise in four data sets. Furthermore, we set the parameters of DP as follows: the sampling rate is 0.01, and the $\delta$ is 1e-3. The result is shown in Table \ref{tab: differential privacy}, we can observe that although the performance of the model is reduced after adding Gaussian noise, it is not much different from the original model. It proves that we can utilize the DP mechanism to protect data privacy further, and there is little change in the model's performance.

\subsubsection{Effect of the Number of Multi-Head Blocks}
In Transformer, the number of multi-head blocks affects the performance of the model. Thus, we experiment with this to find out the suitable number of multi-head blocks. Specifically, the setting of multi-head blocks is related to the original feature dimensions $d$ of the input data and requires that the number of multi-head blocks can be evenly divided by the total number of original feature dimensions $d$. However, we find that when there are more multi-head attention blocks, the model is more prone to overfitting in this experiment. Therefore, we set the number of long blocks on NSL-KDD, Spambase, Shuttle, and Arrhythmia data sets to $2$, $3$, $3$, and $3$, respectively.

\subsection{Discussion}
Here, we analyze in detail the reasons for the above experimental results. First, it can be seen from the above experimental results that our model is better than the baselines in terms of performance and convergence. Specifically, we evaluate the model's performance under different edge device number settings to show that it can be widely used in cloud manufacturing. The experimental results show that FedAnomaly can achieve good scalability. Our framework is an extension of the FL framework, which inherits the good extensibility of the FL framework. In addition, we prove that the proposed framework can reduce communication overhead and we utilize Gaussian noise to ensure the security of the proposed framework. However, one disadvantage of FedAnomaly is its slow convergence. We analyze the reasons for this phenomenon in the following aspects: (1) Our model employs a Transformer that acts as an encoder to extract the data features. However, the effectiveness of the Transformer mainly relies on the self-attention mechanism and multi-head attention block to extract important features. These techniques are too concerned about extracting more important things, which leads to a decrease in convergence speed. (2) Since the distributed training protocol limits the speed of model convergence, the model convergence speed in FedAnomaly is slower than others.

\section{Conclusion}
In this paper, we proposed a new federated learning-based Transformer framework called FedAnomaly to cope with the problems that timely find out the anomalies and protect data privacy in cloud manufacturing. Second, to avoid the conflict between the training of the feature learner and FL training in the proposed framework, we designed a novel cooperative encoder-decoder training protocol to solve the above problems. Compared with data sharing methods and traditional FL solutions, our framework reduces expensive communication overhead while protecting data privacy than data sharing methods and traditional FL solutions. In addition, we verified the effectiveness of our approach by comparing it with centralized methods and advanced anomaly detection methods on four benchmark data sets. In future work, we will explore a framework for real-time detection of data anomalies at each edge device under the premise of protecting data privacy and finding new abnormal situations.

\bibliographystyle{IEEEtran}
\bibliography{ref}

\begin{thebibliography}{10}
\providecommand{\url}[1]{#1}
\csname url@samestyle\endcsname
\providecommand{\newblock}{\relax}
\providecommand{\bibinfo}[2]{#2}
\providecommand{\BIBentrySTDinterwordspacing}{\spaceskip=0pt\relax}
\providecommand{\BIBentryALTinterwordstretchfactor}{4}
\providecommand{\BIBentryALTinterwordspacing}{\spaceskip=\fontdimen2\font plus
\BIBentryALTinterwordstretchfactor\fontdimen3\font minus
  \fontdimen4\font\relax}
\providecommand{\BIBforeignlanguage}[2]{{%
\expandafter\ifx\csname l@#1\endcsname\relax
\typeout{** WARNING: IEEEtran.bst: No hyphenation pattern has been}%
\typeout{** loaded for the language `#1'. Using the pattern for}%
\typeout{** the default language instead.}%
\else
\language=\csname l@#1\endcsname
\fi
#2}}
\providecommand{\BIBdecl}{\relax}
\BIBdecl

\bibitem{liu2020deep}
Y.~Liu, S.~Garg, J.~Nie, Y.~Zhang, Z.~Xiong, J.~Kang, and M.~S. Hossain, ``Deep
  anomaly detection for time-series data in industrial iot: a
  communication-efficient on-device federated learning approach,'' \emph{IEEE
  Internet of Things Journal}, vol.~8, no.~8, pp. 6348--6358, 2020.

\bibitem{wan2017manufacturing}
J.~Wan, S.~Tang, D.~Li, S.~Wang, C.~Liu, H.~Abbas, and A.~V. Vasilakos, ``A
  manufacturing big data solution for active preventive maintenance,''
  \emph{IEEE Transactions on Industrial Informatics}, vol.~13, no.~4, pp.
  2039--2047, 2017.

\bibitem{tao2018digital}
F.~Tao, H.~Zhang, A.~Liu, and A.~Y. Nee, ``Digital twin in industry:
  State-of-the-art,'' \emph{IEEE Transactions on Industrial Informatics},
  vol.~15, no.~4, pp. 2405--2415, 2018.

\bibitem{shi2016edge}
W.~Shi, J.~Cao, Q.~Zhang, Y.~Li, and L.~Xu, ``Edge computing: Vision and
  challenges,'' \emph{IEEE Internet of Things Journal}, vol.~3, no.~5, pp.
  637--646, 2016.

\bibitem{6742605}
F.~Tao, Y.~Zuo, L.~D. Xu, and L.~Zhang, ``Iot-based intelligent perception and
  access of manufacturing resource toward cloud manufacturing,'' \emph{IEEE
  Transactions on Industrial Informatics}, vol.~10, no.~2, pp. 1547--1557,
  2014.

\bibitem{liu2020privacy}
Y.~Liu, J.~James, J.~Kang, D.~Niyato, and S.~Zhang, ``Privacy-preserving
  traffic flow prediction: A federated learning approach,'' \emph{IEEE Internet
  of Things Journal}, vol.~7, no.~8, pp. 7751--7763, 2020.

\bibitem{7883994}
M.~Wollschlaeger, T.~Sauter, and J.~Jasperneite, ``The future of industrial
  communication: Automation networks in the era of the internet of things and
  industry 4.0,'' \emph{IEEE Industrial Electronics Magazine}, vol.~11, no.~1,
  pp. 17--27, 2017.

\bibitem{liu2021towards}
Y.~Liu, R.~Zhao, J.~Kang, A.~Yassine, D.~Niyato, and J.~Peng, ``Towards
  communication-efficient and attack-resistant federated edge learning for
  industrial internet of things,'' \emph{ACM Transactions on Internet
  Technology (TOIT)}, vol.~22, no.~3, pp. 1--22, 2021.

\bibitem{6742575}
F.~Tao, Y.~Cheng, L.~D. Xu, L.~Zhang, and B.~H. Li, ``Cciot-cmfg: Cloud
  computing and internet of things-based cloud manufacturing service system,''
  \emph{IEEE Transactions on Industrial Informatics}, vol.~10, no.~2, pp.
  1435--1442, 2014.

\bibitem{liu2020federated}
Y.~Liu, X.~Yuan, Z.~Xiong, J.~Kang, X.~Wang, and D.~Niyato, ``Federated
  learning for 6g communications: Challenges, methods, and future directions,''
  \emph{China Communications}, vol.~17, no.~9, pp. 105--118, 2020.

\bibitem{hussain2019artificial}
B.~Hussain, Q.~Du, A.~Imran, and M.~A. Imran, ``Artificial intelligence-powered
  mobile edge computing-based anomaly detection in cellular networks,''
  \emph{IEEE Transactions on Industrial Informatics}, vol.~16, no.~8, pp.
  4986--4996, 2019.

\bibitem{7840777}
L.~Stojanovic, M.~Dinic, N.~Stojanovic, and A.~Stojadinovic, ``Big-data-driven
  anomaly detection in industry (4.0): An approach and a case study,'' in
  \emph{Proc. of IEEE Big Data}, 2016.

\bibitem{keshk2019integrated}
M.~Keshk, E.~Sitnikova, N.~Moustafa, J.~Hu, and I.~Khalil, ``An integrated
  framework for privacy-preserving based anomaly detection for cyber-physical
  systems,'' \emph{IEEE Transactions on Sustainable Computing}, vol.~6, no.~1,
  pp. 66--79, 2019.

\bibitem{erfani2014privacy}
S.~M. Erfani, Y.~W. Law, S.~Karunasekera, C.~A. Leckie, and M.~Palaniswami,
  ``Privacy-preserving collaborative anomaly detection for participatory
  sensing,'' in \emph{Proc. of PAKDD}.\hskip 1em plus 0.5em minus 0.4em\relax
  Springer, 2014, pp. 581--593.

\bibitem{vasilomanolakis2015taxonomy}
E.~Vasilomanolakis, S.~Karuppayah, M.~M{\"u}hlh{\"a}user, and M.~Fischer,
  ``Taxonomy and survey of collaborative intrusion detection,'' \emph{ACM
  Computing Surveys (CSUR)}, vol.~47, no.~4, pp. 1--33, 2015.

\bibitem{7430287}
M.~Ozsoy, K.~N. Khasawneh, C.~Donovick, I.~Gorelik, N.~Abu-Ghazaleh, and
  D.~Ponomarev, ``Hardware-based malware detection using low-level
  architectural features,'' \emph{IEEE Transactions on Computers}, vol.~65,
  no.~11, pp. 3332--3344, 2016.

\bibitem{liu2020rc}
Y.~Liu, X.~Yuan, R.~Zhao, Y.~Zheng, and Y.~Zheng, ``Rc-ssfl: Towards robust and
  communication-efficient semi-supervised federated learning system,''
  \emph{arXiv preprint arXiv:2012.04432}, 2020.

\bibitem{9208761}
Y.~Guo, T.~Ji, Q.~Wang, L.~Yu, G.~Min, and P.~Li, ``Unsupervised anomaly
  detection in iot systems for smart cities,'' \emph{IEEE Transactions on
  Network Science and Engineering}, vol.~7, no.~4, pp. 2231--2242, 2020.

\bibitem{voigt2017eu}
P.~Voigt and A.~Von~dem Bussche, ``The eu general data protection regulation
  (gdpr),'' \emph{A Practical Guide, 1st Ed., Cham: Springer International
  Publishing}, vol.~10, p. 3152676, 2017.

\bibitem{dwork2008differential}
C.~Dwork, ``Differential privacy: A survey of results,'' in \emph{Proc. of
  TAMC}.\hskip 1em plus 0.5em minus 0.4em\relax Springer, 2008, pp. 1--19.

\bibitem{9465358}
Y.~Zhou, X.~Song, Y.~Zhang, F.~Liu, C.~Zhu, and L.~Liu, ``Feature encoding with
  autoencoders for weakly supervised anomaly detection,'' \emph{IEEE
  Transactions on Neural Networks and Learning Systems}, pp. 1--12, 2021.

\bibitem{mcmahan2017communication}
B.~McMahan, E.~Moore, D.~Ramage, S.~Hampson, and B.~A. y~Arcas,
  ``Communication-efficient learning of deep networks from decentralized
  data,'' in \emph{Proc. of AISTATS}, 2017.

\bibitem{zheng2022aggregation}
Y.~Zheng, S.~Lai, Y.~Liu, X.~Yuan, X.~Yi, and C.~Wang, ``Aggregation service
  for federated learning: An efficient, secure, and more resilient
  realization,'' \emph{IEEE Transactions on Dependable and Secure Computing},
  2022.

\bibitem{liu2012unsupervised}
Z.~Liu, R.~Shi, L.~Shen, Y.~Xue, K.~N. Ngan, and Z.~Zhang, ``Unsupervised
  salient object segmentation based on kernel density estimation and two-phase
  graph cut,'' \emph{IEEE Transactions on Multimedia}, vol.~14, no.~4, pp.
  1275--1289, 2012.

\bibitem{8478396}
N.~Lv, C.~Chen, T.~Qiu, and A.~K. Sangaiah, ``Deep learning and superpixel
  feature extraction based on contractive autoencoder for change detection in
  sar images,'' \emph{IEEE Transactions on Industrial Informatics}, vol.~14,
  no.~12, pp. 5530--5538, 2018.

\bibitem{7539352}
K.~Chen, J.~Hu, and J.~He, ``Detection and classification of transmission line
  faults based on unsupervised feature learning and convolutional sparse
  autoencoder,'' \emph{IEEE Transactions on Smart Grid}, vol.~9, no.~3, pp.
  1748--1758, 2018.

\bibitem{5356528}
M.~Tavallaee, E.~Bagheri, W.~Lu, and A.~A. Ghorbani, ``A detailed analysis of
  the kdd cup 99 data set,'' in \emph{2009 IEEE Symposium on Computational
  Intelligence for Security and Defense Applications}, 2009, pp. 1--6.

\end{thebibliography}

\end{document}